# Quantitative Characterization of Combinatorial Transcriptional Control of the Lactose Operon of *E. coli*

Thomas Kuhlman[§], Zhongge Zhang[¶], Milton H. Saier, Jr.[¶], and Terence Hwa[§*]

[§]*Center for Theoretical Biological Physics* and [¶]*Division of Biological Sciences,*

*University of California at San Diego, La Jolla, CA  92093-0374*

[*]To whom correspondence should be addressed. *e-mail: hwa@ucsd.edu*

**Corresponding Author** Terence Hwa,
Center for Theoretical Biological Physics,
University of California at San Diego,
9500 Gilman Drive, La Jolla, CA 92093-0374
phone: 858 534-7263; fax: 858-534-7697;
email: hwa@ucsd.edu

**Abbreviations**  LacR: the Lac repressor; CRP: cAMP-receptor protein.

# ABSTRACT

It is the goal of systems biology to understand the behavior of the whole in terms of the knowledge of the parts. This is hard to achieve in many cases due to the difficulty of characterizing the many constituents and their complex web of interactions involved in a biological system. The *lac* promoter of *E. coli*, being one of the most extensively studied systems of molecular biology, offers a possibility of confronting "system-level" properties of transcriptional regulation with the known biochemistry of the molecular constituents and their mutual interactions. Such confrontations can reveal previously unknown constituents and interactions, as well as offering new insight into how the components work together as a whole. Here we study the combinatorial control of the *lac* promoter by the regulators LacR and CRP. A previous *in vivo* study [Setty et al, PNAS 100: 7702-7 (2003)] found gross disagreement between the observed promoter activities and the expected behavior based on the known molecular mechanisms. We repeated the study by identifying and removing several extraneous factors which significantly modulated the expression of the *lac* promoter. Through quantitative, systematic characterization of promoter activity for a number of key mutants and guided by the thermodynamic model of transcriptional regulation, we are able to account for the combinatorial control of the *lac* promoter quantitatively, in term of a cooperative interaction between CRP and LacR-mediated DNA looping. Specifically, our analysis indicates that the sensitivity of the inducer response results from LacR-mediated DNA looping, which is significantly enhanced by CRP.

# INTRODUCTION

The *lac* promoter of *E. coli* is one of the most extensively studied systems of molecular biology (1-6). Knowledge and insight gained from these studies have shaped much of how we now think about gene regulation. It is well known that *E. coli* cells repress the expression of the *lac* operon when glucose is abundant in the growth medium. Only when the glucose level is low *and* the lactose level is high is the operon fully expressed. Thus the regulation of this operon represents an example of "combinatorial control" widely seen in prokaryotes and eukaryotes (7, 8). In this case, the combinatorial control is implemented molecularly by two transcription factors, the Lac repressor LacR which represses transcription and the cyclic-AMP receptor protein CRP which activates transcription. Activation by CRP requires the inducer cAMP, which is used by *E. coli* cells as a signal of glucose shortage (9-15). Repression by LacR is activated in a nearly all-or-none manner upon varying the amount of lactose or one of the several synthetic inducers in growth medium with poor carbon sources (16, 17). The fold-change in repression is very large, (> 1000-fold), and has been shown to involve LacR-mediated DNA looping (5, 6, 18-21).

Here we investigate quantitatively the competing effects of activation and repression on the *lac* promoter *in vivo*. We focus on two perplexing issues: (i) According to biochemical studies (22-25), LacR-inducer interaction is only weakly cooperative. By what mechanism(s) does the observed induction response become so abrupt (26, 27)? (ii) Despite the well-known role CRP plays in activating transcription (28), structural studies (29, 30) suggest that CRP enhances *repression* by facilitating the LacR-mediated DNA looping. Moreover, *in vitro* biochemical studies indicate that CRP stabilizes LacR-DNA binding (31-33). What functional role(s) does CRP actually play in the control of this operon? We approached these issues by first identifying mutants of *E. coli* MG1655 that allowed us to directly control the activities of the activators and repressors by varying the levels of two inducers in the growth medium. We then characterized the promoter activity systematically for numerous combinations of the inducers. The gene expression data obtained clearly reveal the effect of CRP in enhancing the steepness of the inducer response.  We developed a thermodynamic model of gene regulation (8, 34-36), incorporating the known molecular mechanisms of LacR-induced DNA looping and its coupling to CRP through DNA bending (32, 33). The success of the model is manifested in its ability to describe the complex co-dependence of gene expression on the two inducer levels quantitatively by invoking a *single* parameter, the cooperativity between CRP and LacR-mediated DNA looping, with the fitted value of the cooperativity agreeing well with that determined from *in vitro* biochemical measurements (32, 33). Our study presents a proof of concept that the complicated web of interactions coupling repressors, activators, promoters, and DNA loops *in vivo* can be quantitatively dissected, provided that the right modeling, together with a precise sequence of experiments on a systematically picked set of mutants are carried out.

# RESULTS

## Repression by LacR

We first characterized quantitatively the activity of the *lac* promoter (P*lac*) subject to varying degrees of repression by the Lac repressor, LacR, for *E. coli* cells in the exponential growth phase. In our experiments, the activity of LacR was modulated by the synthetic gratuitous inducer isopropyl β-D-thiogalactopyranoside (IPTG) (37). We performed the β-galactosidase assays for wildtype *E. coli* MG1655 cells in M9 minimal medium with 0.5% glucose as the carbon source and up to 1 mM of IPTG. "P*lac* activity", defined here as the *product* of the β-galactosidase activity and the cell-doubling rate (see Supp Mat), is plotted against the corresponding IPTG concentrations as the black crosses in Fig. 1a.

The data points are fitted to the Hill function

$$\alpha_{IPTG} = b_{IPTG} \cdot \frac{1 + f_{IPTG} \cdot \left([IPTG]/C_{IPTG}\right)^{m_{IPTG}}}{1 + \left([IPTG]/C_{IPTG}\right)^{m_{IPTG}}}, \tag{1}$$

shown as the black line in Fig. 1a. The Hill function is used here (and elsewhere in the Results section) merely to extract the *qualitative* features of the promoter activity; the appropriate quantitative description of the data presented will be provided below in the Analysis section. The features of the IPTG-dependent promoter activity (the "IPTG response") are conveyed by the best-fit Hill parameters listed in Table 1 (row 1): The overall fold change, quantified by $f_{\mathrm{IPTG}}$ ~ 1200, is in good agreement with previous studies (5, 6, 18-20, 38, 39). However, the "sensitivity" of the response, quantified by the slope of the transition region in a log-log plot and given approximately by the Hill coefficient ($m_{IPTG} \approx 4.5$), is much larger than the expected behavior based on the known biochemistry; see below.

One factor contributing to the hyper-sensitivity observed is a positive-feedback effect due to the expression of *lacY* which encodes the Lac permease. This effect is analogous to the well known effect on the induction of the *lac* promoter due to LacY expression (16). Although the uptake of IPTG does not require LacY, Jensen et al (40) showed that the expression of LacY by the *lac* promoter nevertheless led to enhanced sensitivity. We therefore deleted *lacY* from *E. coli* MG1655 to form strain TK150 (see Table S1), and repeated the β-galactosidase assays for this strain. P*lac* activities obtained are plotted as the cyan crosses in Fig. 1a. Fitting again to the Hill form yields the cyan line with the best-fit parameters shown in Table 1 (row 2). The IPTG response of the Δ*lacY* mutant exhibits a broader transition, with $m_{IPTG} \approx 2.6$.

Another possible cause of this hyper-sensitivity is a suggested cooperative interaction between the Lac repressor and the activator CRP which also binds in the promoter region (31-33). To investigate this possibility, we deleted the *crp* gene from *E. coli* TK150 to form strain TK230 (Table S1), and repeated the β-galactosidase assays for this strain. P*lac* activities obtained are plotted as the green circles in Fig. 1a. Fitting again to the Hill form yields the green line with the best-fit parameters shown in Table 1 (row 3). The IPTG response of the Δ*crp* Δ*lacY* double mutant is broader still, with $m_{IPTG} \approx 2$.

## Activation by CRP

We next characterized the dependence of the P*lac* activity to different degrees of activation by cAMP-CRP (the "cAMP response"). To avoid possible complications due to interaction between LacR and CRP, all experiments were performed under saturating IPTG concentration (1 mM) to disable LacR-operator interaction. This was complemented by direct deletion of LacR in some cases (see below). We also deleted *lacY* in all subsequent experiments in order to avoid possible feedback. In all cases discussed below where we directly compared the P*lac* activity of $lacY^-$ and $lacY^+$ cells, differences of no more than 2-fold were obtained (data not shown).

**Control by cAMP**. One way to manipulate the cellular level of cAMP-CRP is to subject cells to different levels of cAMP in the medium, and rely on the diffusion of cAMP into cells. This approach requires shutting off the endogenous synthesis of cAMP by the enzyme adenylate cyclase (AC), encoded by *cyaA* (41-44). Setty et al (27) attempted this approach by growing cells in medium with 0.2% glucose and varying levels of cAMP, expecting that AC activity would be repressed via catabolite repression (10, 14, 15, 45-49). However, they observed only a few-fold change in P*lac* activity despite large variations in the extracellular cAMP levels (0 to 20 mM) (27). The observed change was surprisingly small given that > 50-fold difference in P*lac* activity was obtained between the wildtype and Δ*crp* strains (3). In fact, a nearly 10-fold difference in P*lac* activity can be seen by simply growing wildtype cells on various sugars; see Fig. S1 (blue bars).

**Effect of Adenylate Cyclase deletion.** We reasoned that the small change in P*lac* activity obtained by Setty et al might have resulted from the incomplete repression of AC activity by glucose uptake, and hence repeated the experiment with the deletion of *cyaA*. *E. coli* TK250 strain (Δ*cyaA* Δ*lacY*, see Table S1) was grown in M9 minimal medium with 0.5% glucose, 1mM IPTG, and up to 10mM cAMP. β-galactosidase activity was assayed as described above. The resulting P*lac* activity displays a smooth sigmoidal dependence as shown in Fig. 1b (blue squares). [Almost identical results (not shown) were obtained for *LacR* null mutant (strain TK320), indicating that LacR is indeed not a factor with saturating IPTG (1 mM) in the growth medium.] A ~100-fold difference is seen between the low and high cAMP concentrations, comparable to the difference in P*lac* activity reported between *crp* null mutant and wildtype strains of *E. coli* (3). In contrast, P*lac* activity of the wildtype (black symbols in Fig. 1b) as well as the Δ*lacY* mutant (not shown) grown in glucose displayed only ~ 3-fold change over the same range of cAMP levels, similar to the afore-mentioned finding by Setty et al (27). As a negative control, we show in Fig. 1b (green circles) the promoter activity obtained for the *crp*$^-$ strain (TK230); its lack of cAMP dependence indicates that the observed cAMP dependence for the *crp*$^+$ strain was mediated primarily by cAMP-CRP. Additional negative controls on possible indirect effects of cAMP variations on P*lac* activity are shown in Fig. S2b. We found the variation of CRP expression and the non-specific effects of CRP on P*lac* activity to be small (~2-fold), compared to the 100-fold difference observed for the Δ*cyaA* mutants over the same range of cAMP concentrations.

The cAMP response exhibited by strain TK250 was analyzed by fitting to the Hill function,

$$\alpha_{cAMP} = b_0 \cdot \frac{1 + f_{cAMP} \cdot \left([cAMP]/C_{cAMP}\right)^{m_{cAMP}}}{1 + \left([cAMP]/C_{cAMP}\right)^{m_{cAMP}}}, \tag{2}$$

and plotted as the blue line in Fig. 1b. The best-fit parameters are shown in Table 2 (row 1). The sensitivity of the cAMP response, characterized by the Hill coefficient $m_{\mathrm{cAMP}} \approx 2$, is in disagreement with the non-cooperative nature of cAMP-CRP interaction (50-53). This suggests a nonlinear relationship between the extracellular and intracellular cAMP concentrations in Δ*cyaA* strain, and prompted us to look for additional factors regulating intracellular cAMP levels.

**Effect of Phosphodiesterase deletion**. One such factor is cAMP degradation catalyzed by the enzyme cAMP-phosphodiesterase (PDE) (54-58), encoded by *cpdA* (59). We deleted the *cpdA* gene to obtain strain TK310 (Δ*cyaA* Δ*lacY* Δ*cpdA*) (see Table S1) and repeated the β-galactosidase assay and analysis. The cAMP response obtained (red circles in Fig. 1b) is more gradual than that of $cpdA^+$ cells (blue squares). Fitting the data to the Hill form (2) yielded the red line with the Hill parameters listed in Table 2 (row 2). Specifically, the Hill coefficient $m_{\mathrm{cAMP}} \approx 1$ is consistent with the naïve expectation based on the non-cooperative nature of cAMP-CRP interaction (50-53).

## Combinatorial Control

We next investigated the *co-dependence* of P*lac* activity on the two regulators, LacR and cAMP-CRP. We showed above that the Δ*crp* Δ*lacY* strain (TK230) could be used to characterize the bare IPTG response, while the Δ*lacY* Δ*cyaA* Δ*cpdA* strain (TK310) could be used to characterize the bare cAMP response. In order to characterize the co-dependence of the promoter on IPTG and cAMP, we first verified that strain TK310 exhibited nearly the same IPTG dependence as TK230 in growth medium with no cAMP added; compare[1] the red squares and green circles in Fig. 1a. Fitting the IPTG-response of TK310 (red squares) to the Hill form (1) yields the solid red line; the corresponding parameters are provided in Table 1 (row 4).

We repeated the β-galactosidase assay and analysis for TK310 cells grown in media with various *combinations* of IPTG and cAMP concentrations. As evidence of interaction between IPTG-mediated and cAMP-mediated regulations, we show in Fig. 1a (red triangles) the IPTG response for TK310 strain in growth medium containing 1mM cAMP. This response is nearly indistinguishable from the physiological P*lac* activity exhibited by *cyaA*$^+$ *cpdA*$^+$ cells in glycerol medium (cyan circles). The dashed red line is the best fit to the Hill function (1), with parameters listed in Table 1 (row 5). Note that the overall fold-change ($f_{\mathrm{IPTG}}$) is increased from < 250x when no cAMP was added in the medium, to > 1500x with 1mM cAMP in the medium. The latter fold-change is comparable to those obtained for wildtype *E. coli* cells grown in the absence or saturating concentration of IPTG (black crosses in Fig. 1a, and Refs. (5, 6, 18-20, 38, 39)). Additionally, the sensitivity of the IPTG response increased from $m_{\mathrm{IPTG}} \approx 2.0$ (Table 1 rows 4) for TK310 cells grown in the absence of cAMP to $m_{\mathrm{IPTG}} \approx 2.8$ (Table 1 rows 5) for the same cells grown in 1mM cAMP. Fitting the IPTG responses of these cells obtained at a variety of cAMP concentrations, we found a trend of increasing Hill coefficient (from 2 to 3) and fold-change (from 250x to 1800x) for cAMP levels from 1 μM to 1mM; see Figs. S3. The complete co-dependence of P*lac* activity on IPTG and cAMP is shown as the 3D plot in Fig. S4a.

[1] However, *cpdA* mutants show a 2-fold overall reduction in gene expression for unknown reasons.

# ANALYSIS

We have seen that the IPTG and cAMP responses of various mutant strains of *E. coli* MG1655 fitted well to Hill functions, with the Hill parameters summarized in Tables 1 and 2. However, the Hill function itself has been invoked so far without justification; it was merely a familiar form used to quantify key features of the response, e.g., the overall fold-change and sensitivity. Below we will analyze and interpret the results obtained in light of the rich knowledge on the molecular biology of the *lac* promoter and the biochemistry of the associated components using a thermodynamic model of transcriptional regulation; see Refs. (8, 36) and a brief review in Supp. Mat.

**Activation by cAMP-CRP is non-cooperative.** The cAMP response found for strain TK310 (Δ*cyaA* Δ*cpdA* Δ*lacY*) exhibited a broad transition (red circles in Fig. 1b), well fitted by the Hill form (red line). This is in agreement with the thermodynamic model which predicts the Hill form for response to simple activation by cAMP-CRP; see Bintu et al (35) and Supp. Mat. The Hill coefficient $m_{cAMP} \approx 1$ obtained is in good agreement with the biochemistry finding that it takes one cAMP molecule to activate the CRP dimer (51-53).

The thermodynamic model further relates the other Hill parameters $f_{cAMP}$ and $C_{cAMP}$ to the biochemical parameters that describe CRP-mediated transcriptional activation; see Supp. Mat. Best estimates of $f_{cAMP}$ and $C_{cAMP}$ based on knowledge of the biochemical parameters are given in Table 2 (row 3). The available information is not sufficient for a quantitative comparison of the parameter $C_{cAMP}$, whose value depends on the CRP-operator binding affinity *in vivo* as well as the relation between the intra- and extra- cellular cAMP concentrations, both of which can be estimated only very crudely (9, 60); see Supp. Mat. The parameter $f_{cAMP}$, which describes the maximal fold-change in the cAMP response, is given by the cooperativity of CRP and RNA polymerase (RNAp) interaction according to the thermodynamic model. Our result $f_{cAMP} = 240 \pm 13$ is significantly larger than the cooperativity factor of ~20 obtained from *in vitro* biochemical measurements (61, 62). This discrepancy is analogous to one noted earlier by Beckwith et al (3). It may have resulted from the accumulation of a number of small factors. For example, in addition to recruiting RNAp, CRP was shown to stimulate the transition of promoter DNA from the close to open conformation, thereby enhancing the transcription rate by ~50% (61). Also, the autoregulation of CRP expression may account for ~2-fold difference (see Fig. S2b and caption), and another 2~3 fold difference may be attributed to the deletion of *cpdA* (compare the vertical ranges of the blue and red lines in Fig. 1b).

**PDE provides insulation to variations in cAMP.** We are not aware of any significant phenotype reported for cells with a PDE deletion. Only small differences of 2 – 3 fold in P*lac* activity were seen between the wildtype and Δ*cpdA* strains (blue and red bars in Fig. S1), while no systematic trend can be seen in the growth rates of the two strains (the numbers on top of the bars in Fig. S1) either. However, the effect of *cpdA* expression on the *cyaA* mutant is striking. Comparison of the blue and red lines in Fig. 1b suggests that PDE expression *insulates* the cell from extracellular cAMP variations of up to 100 μM. This may be important for cells in environments where AC activity is significantly repressed.

**LacR-mediated DNA looping increases the sensitivity of the IPTG response**. The IPTG response of the Δ*crp* Δ*lacY* double mutant (TK230) exhibited a reduced sensitivity (Fig. 1a green circles) compared to the *lacY* mutant (TK150, cyan crosses in Fig. 1a). The difference is due to the activated CRP in the latter strain (which has *cyaA* intact and hence synthesizes cAMP endogenously). We will discuss the effect of CRP shortly; for now, we first discuss the IPTG response in the absence of CRP, i.e., that of strain TK230. This response is well fitted by the Hill form (green line), with the Hill coefficient $m_{\mathrm{IPTG}} \approx 2.0$ (Table 1 row 3). While a cooperative IPTG response with Hill coefficient ~2 is widely quoted in the molecular biology literature (26), and moreover a cooperative IPTG-LacR interaction was suggested based on a structural study of LacR (29), *in vitro* biochemical studies of IPTG-LacR binding in fact found a non-cooperative IPTG-LacR interaction, which became only weakly cooperative (Hill coefficient = 1.4 ~1.6) in the presence of operator DNA fragments (22-25). We reasoned that the apparent cooperativity ($m_{\mathrm{IPTG}} \sim 2$) observed *in vivo* might have resulted from LacR-mediated *DNA looping*[2] known to occur in wildtype *E. coli* cells (6, 18-20). The corresponding thermodynamic model was developed in Supp Mat, with the following form of the IPTG response

$$\alpha_{IPTG} = \alpha_0 \Big/ \left[ 1 + \frac{R}{\left(1 + [\mathrm{IPTG}]/K_{\mathrm{IPTG}}\right)^2} + \frac{R \cdot L_0}{\left(1 + [\mathrm{IPTG}]/K_{\mathrm{IPTG}}\right)^4} \right]. \tag{3}$$

Although Eq. (3) is not of the Hill form, it resembles a spectrum of different Hill functions, with the maximal fold-change

$$f_{IPTG} = 1 + R \cdot (L_0 + 1), \tag{4}$$

and with the sensitivity also increasing with the parameters $R$ and $L_0$; see Supp Mat. The parameter $R$ depends on the LacR concentration and LacR-operator affinity; it gives the fold-repression between zero and saturating IPTG levels in the absence of DNA looping ($L_0 = 0$). The values of $R \approx 20$ - $50$ can be inferred from two previous studies by Oehler et al (6, 20), who characterized the activities of *lac* promoters bearing various Lac operator mutations; see Supp Mat. In the analysis below, we will use $R \approx 50$; results of similar quality (not shown) can be obtained for $R \approx 20$. The parameter $L_0$ describes the enhancement in the "local concentration" of LacR molecules due to DNA looping (6, 35). There is currently no direct measurement of $L_0$ for the *lac* promoter in the absence of CRP binding.

---

[2] We note that the increased in cooperativity by DNA looping was discussed recently by Vilar & Saiz (75) in the context of the lysis-lysogeny control of Phage λ. However, the mechanism of cooperativity there is the oligomerization of regulatory proteins made possible through DNA looping. In contrast here, the tetramerization of LacR is independent of DNA looping. What DNA looping enhances here is the *sensitivity* to IPTG rather than LacR.

Before applying Eq. (3) to our results, we first discuss its form in the absence of DNA looping ($L_0 = 0$). In a very recent report, Oehler et al (63) used various mutants to characterize the IPTG responses of the *lac* promoter incapable of DNA looping. Their data are reproduced in Fig. S5(a) (red triangles and blue squares) and shown to fit well to Eq. (3) with $L_0 = 0$ (solid line). The data are equally well-fitted in Fig. S5(b) by the Hill form (1) with an effective Hill coefficient of 1.5, in agreement with results of previous *in vitro* studies (24). Together, the goodness of these fits validates the above quantitative description of LacR-operator interaction based on the thermodynamic model.

We next confront DNA looping by fitting the IPTG response of strain TK230 (the green circles in Fig. 1(a)) to Eq. (3), with two fitting parameters, $K_{IPTG}$ and $L_0$. (The saturation value $\alpha_0$ is fixed by the promoter activity at 1mM IPTG.) The best-fit curve (not shown) is nearly indistinguishable from the green line in Fig. 1a even though Eq. (3) contains one fewer fitting parameter than Eq. (1). In particular, the observed sensitivity of the IPTG dependence is well reproduced. The best-fit parameters obtained are shown in Table 3 (row 1). The value of the loop parameter simply reflects the maximal fold-change of the IPTG response as given by Eq. (4). The value of the IPTG-LacR dissociation constant $K_{IPTG}$ obtained is comparable to the results of *in vitro* biochemical studies; see Refs. (4, 64, 65) and Supp Mat. Similarly, we found that the DNA looping model described the IPTG response of strain TK310 (Δ*cyaA* Δ*cpdA* Δ*lacY*) very well in medium with no cAMP. The corresponding best-fit parameters are given in Table 3 row 2; the best-fit curve (not shown) is nearly indistinguishable from the lower red line in Fig. 1a.

**Cooperativity between CRP and LacR-mediated DNA looping enhances the sensitivity of the IPTG response.** Strain TK310, which exhibited the expected cAMP response in medium with 1mM IPTG, and the expected IPTG response in medium with no cAMP added, was used to investigate the *combinatorial control* of P*lac* activity by LacR and cAMP-CRP. According to the thermodynamic model, if there is no interaction between cAMP-CRP mediated activation and LacR mediated repression, then the co-dependence of P*lac* activity on IPTG and cAMP would simply be the algebraic *product* of the IPTG response and the cAMP response, i.e., the product of the red lines in Fig. 1a and Fig. 1b. A manifestation of this product form is a simple vertical shift of the IPTG responses of the same strain at different cAMP concentrations in the log-log plots. However, this clearly cannot be the case for strain TK310, as both the sensitivity and the maximal fold-change of its IPTG response increase with increasing cAMP levels (Fig. 1a and Fig. S3); see also Fig. S8(a) for a direct comparison.

Various forms of interactions between CRP and LacR in the *lac* promoter region have been previously proposed or reported (29, 31-33, 66, 67): Since CRP causes a $90^o$-$130^o$ bend in DNA (20, 68, 69), the binding of CRP to its site in the *lac* promoter could potentially bring the Lac operators O1 and O3 into closer proximity, promoting the formation of DNA loop between them (29); see Fig. S6. The existence of this interaction is supported by a computational study based on the available LacR-operator and CRP-operator co-crystal structures (30). Moreover, Hudson & Fried (32) and Vossen et al (33) found cooperativity in the formation of a ternary complex of CRP, LacR, and *lac* promoter DNA *in vitro*, with the simultaneous presence of LacR and cAMP-CRP increasing the affinity of both proteins for their respective binding sites by a cooperative factor (Ω) of 4-12 fold.

To see whether CRP-assisted DNA looping may *quantitatively* explain the observed co-dependency of P*lac* activity, we extended the thermodynamic model with DNA looping to include the CRP-dependent looping effect; see Supp Mat. The extended model predicts the IPTG response to have the same form as Eq. (3), but with the prefactor $\alpha_0$ replaced by the Hill function $\alpha_{\mathrm{cAMP}}$ for strain TK310 (Eq. (2) with parameters given by Table 2 row 2), and with the loop parameter $L_0$ replaced by

$$L = L_0 \cdot \frac{1+\Omega\cdot[\mathrm{cAMP}]/C_{cAMP}}{1+[\mathrm{cAMP}]/C_{cAMP}} \tag{5}$$

where Ω is the cooperativity factor of the CRP-LacR-loop interaction.

To test the predictions of the extended thermodynamic model, we used Eq. (4) to infer the value of the loop parameter $L$ from the maximal fold-change of the IPTG response at each cAMP value; the result is plotted as the circles in Fig. 2a. We then used Eq. (3) to compute the expected IPTG responses using these values of $L$. The results are shown as the dashed lines in Fig. 2b for several representative cAMP values, together with the experimental data for TK310 cells grown in media with the corresponding cAMP concentrations. Note that the dashed lines are *not* fits of the data to the model. The striking congruence between the dashed lines and the data points (for $[\mathrm{cAMP}] \leq 1$ mM) shows that the cAMP-dependent sensitivity, which requires a series of different Hill coefficients within the Hill function description (Fig. S3a), can be naturally described by the DNA looping model in terms of the cAMP-dependent loop parameter $L$ with *no* other adjustable parameters.

To further examine whether the cAMP-dependence of $L$ conforms to the known energetics of CRP-assisted DNA looping, we fitted the inferred values of $L$ (Fig. 2a circles) to Eq. (5), by allowing only the parameter $\Omega$ to vary. The result of the fit (Fig. 2a solid line) described the data well, except again at the highest cAMP level (10 mM) where cell growth significantly slowed down and other effects (e.g., over expression of nucleoid proteins) might significantly affect DNA bending and looping (70). The value of the best-fit parameter obtained, $\Omega = 10.3 \pm 0.1$, is in agreement with the 4-12 fold cooperativity found *in vitro* (31, 33); see Supp Mat. Fig. S4b shows the 3D plot of the expected P*lac* activity according to the extended thermodynamic model. Collectively, these results provide strong quantitative support for the hypothesis of CRP-assisted repression by DNA looping (31-33).

Finally, we repeated the analysis for the combinatorial control of TK250 cells which are *cpdA*$^+$. Despite the unexpected cooperative nature of the cAMP response of this strain (blue symbols in Fig. 1b with $m_{cAMP} \approx 2$), the IPTG responses at various cAMP levels (Fig. S7b) are still well described by Eq. (3) (with the best-fit values of $K_{\mathrm{IPTG}}$ and $L_0$ given in Table 3 row 3). Moreover, except for the highest cAMP concentrations again, the cAMP-dependence of the loop parameter $L$ agrees well with the corresponding Hill function, with similar value of the cooperativity factor, $\Omega \approx 9.7$; see Fig. S7a and caption. These results indicate that *cpdA* expression only affects the activation of CRP by extracellular cAMP, but not the interaction of cAMP-CRP with DNA looping itself.

## DISCUSSION

Systematic, quantitative characterization of gene expression may yield unique insights on the mechanisms of gene regulation. A major obstacle in this effort is to quantify the amount of *active* transcription factors in cells in a given environment. As attempted earlier by Setty et al (27), we tackled this problem for the *lac* promoter by controlling the amount of inducers IPTG and cAMP extracellularly, but using *cyaA* mutant of *E. coli* which cannot synthesize cAMP endogenously.

One danger of this approach is the array of mechanisms the cell may employ to control the intracellular effector levels. We encountered two such mechanisms in our study: Through the positive feedback of the Lac permease, *E. coli* cells sensitize themselves to the presence of inducers of the Lac repressor, while through the activity of cAMP-phosphodiesterase (PDE), *E. coli* cells appear to insulate themselves from low levels of cAMP in the environment. The LacY effect is well known for induction with lactose in the growth medium (16, 17). However, the PDE effect observed here is previously unknown. It may yield important clues towards a comprehensive understanding of cAMP transport and control central to carbon metabolism in *E. coli*.

**DNA looping is assisted by CRP**. The innate regulation of P*lac* activity is revealed in *E. coli* cells bearing *cyaA lacY cpdA* triple deletion. We found that with a saturating amount of IPTG in the medium, the cAMP-CRP-mediated activation behaved in accordance with the thermodynamic model, given the non-cooperative nature of the cAMP-CRP interaction. Also, in the absence of cAMP, the LacR-mediated repression behaved in accordance with the thermodynamic model with DNA looping. Moreover, the co-dependence on IPTG and cAMP can be explained by the extended thermodynamic model with a single cooperativity parameter $\Omega$ whose value is of the order found by *in vitro* experiments. Both the quality of the fit and the simplicity of the model (i.e., one parameter for co-dependence) are superior compared to other simple but *ad-hoc* models, as we show explicitly in Supp. Mat. (Figs. S8). These results strongly support the hypothesis of CRP-assisted DNA looping (29, 31), and also shed light on the seemingly conflicting roles CRP plays in the regulation of the *lac* promoter: CRP is an activator in that it increases gene expression at each IPTG level. However it increases gene expression in a nonlinear way such that the maximal fold-change and the sensitivity of the IPTG response are both enhanced.

**Hyper-sensitivity of IPTG response resulted from multiple factors.** Even in the absence of *lacY* mediated feedback, P*lac* activity depended sensitively on the IPTG level with apparent Hill coefficient of $2.5 \sim 3$ at physiological intracellular cAMP concentrations. This is rather remarkable given that the basic interaction of IPTG with the Lac tetramer is not cooperative at all in the absence of operator DNA (24). While the involvement of DNA looping in the IPTG hyper-sensitivity is shown qualitatively by the experiment of Oehler et al (63), the results and analysis of this study reveal the complex way in which this hyper-sensitivity is attained: First, the non-cooperative IPTG-LacR binding becomes weakly cooperative (effective Hill coefficient ~ 1.5) in the presence of operator DNA [Ref. (22-25) and Fig. S5], due presumably to the allosteric coupling between inducer and operator binding (see Supp Mat). The involvement of DNA looping then increases the cooperativity (effective Hill coeff ~ 2), as both dimeric units of the Lac tetramer are required to bind specifically to operators in order to form DNA loops. Finally, in the physiological range of intracellular cAMP concentrations where CRP is activated, the cooperative interplay between DNA looping and CRP-mediated DNA bending further enhances the looping-induced cooperativity to the observed effective Hill coefficient of 2.5 ~ 3.

As shown by Novick & Weiner and by Ozbudak et al (16, 17), the hyper-sensitivity of P*lac* activity to the inducer of LacR is crucial for the bistable (i.e., the all-or-none) nature of *E. coli*'s lactose utilization strategy (16, 17). The latter is the molecular basis of important physiological effects such as the diauxie shift (71). It is interesting that hyper-sensitivity resulted in this case from an agglomeration of weak interactions. This *distributed* way of implementing an important molecular function (instead of, e.g., a highly cooperative LacR-inducer interaction) may reflect the serendipity of the evolutionary dynamics, or alternatively a robust evolutionary strategy to preserve important system-level functions.

**The importance of being quantitative.** Critical to this study was the recognition of various "artifacts" produced by processes not directly related to the regulation of the *lac* promoter. The key discriminating feature we used was the quantitative comparison of the observed responses to the expectations of the thermodynamic model, in light of the known biochemical processes. Although fitting the experimental data to a Hill form is in itself not a very discriminating task, we find that the "reasonableness" of the numerical values of the fitting parameters, especially the apparent Hill coefficients of the responses, can be quite revealing. Nevertheless, the power of the quantitative analysis is by itself limited, as it can only suggest the existence of problems but does not identify the sources. It is through the quantitative comparison of the characteristics of a series of key mutants that the major conclusions of this study are established. It is important to note that some of our most discriminating mutants did not display much difference in the high/low states of expression, but exhibited clear differences in the abruptness of the transition between states.

The *lac* promoter is one of the prototypical model systems of gene regulation. Classic studies on the regulation of this promoter have established numerous fundamental concepts as well as laying down the appropriate methodology for studying the molecular biology of gene regulation. Quantitative studies of the *lac* promoter can again play important roles in laying down the foundation of quantitative systems biology, whose goal is to understand the behavior of a "system" in terms of the relevant properties of its components. The *lac* promoter is admittedly a rather simple system. Nevertheless, we see from this study that system-level properties such as the sensitivity of the IPTG response resulted from a closely intertwined set of interactions among the molecular constituents. We demonstrated how this system can be dissected by careful quantitative characterization and targeted genetic manipulations, along with guidance from quantitative modeling and the knowledge of the biochemistry of the molecular constituents. The experience gained here may be of value to the study of other more complex biological systems.

# METHODS

**Plasmids and Strains.** All strains used in this study were derived from *E. coli* K-12 MG1655 as listed in Table S1 and detailed in Supp Mat. Chromosomal gene deletion were performed using the method of Datsenko & Wanner (72), and transferred from one strain to another using P1 transduction. All mutations were verified using PCR.

**Cell growth and β-galactosidase Assay.** Overnight cultures were grown in M9 minimal medium containing the standard concentrations of necessary antibiotics in a 37$^{o}$C water bath until stationary phase. The carbon source was 0.5% glucose unless otherwise indicated. These cultures were diluted 100- to 1000- fold into 24-well plates (Costar) containing the same growth medium plus varying concentrations of IPTG and cAMP. The plates were grown with vigorous shaking in a humidity-controlled incubator maintained at 37$^{o}$C, with $OD_{600}$ measurements taken every 2 hours in a Tecan Genios Pro plate reader. When $OD_{600}$ of a sample reached 0.2-0.4, it was assayed for β-galactosidase activity. These assays were performed in triplicate or more according to Miller (73) and Griffith (74), with minor modifications detailed in Supp Mat. The β-galactosidase activity obtained (*A*) was expressed in Miller Units (73), and the promoter activity ($\alpha$) reported was taken to be the *product* of *A* and the cell-doubling rate $\lambda_{\frac{1}{2}}$. Serial dilution experiments were used to verify that the entire range of promoter activity reported in the figures lay within the linear responsive regime of the measurements; see Supp Mat.

## ACKNOWLEDGEMENT

We are indebted to Bob Schleif, Bill Loomis and Shumo Liu for pointing out conceptual flaws in early versions of our experiments. We also benefited from valuable comments from many other colleagues throughout the course of this work, including Nicolas Buchler, Antoine Danchin, Hernan Garcia, Kathleen Matthews, Stefan Oehler, Rob Philips, and Agnes Ullmann. This research was supported by the DOE (Grant No. DE-FG02-03-ER63691), the NSF (Grant No. MCB-0417721), and by the NIH (Grant No. GM070915). TK and TH additionally acknowledge support through the NSF PFC-sponsored Center for Theoretical Biological Physics (Grants No. PHY-0216576 and PHY-0225630).

# FIGURES

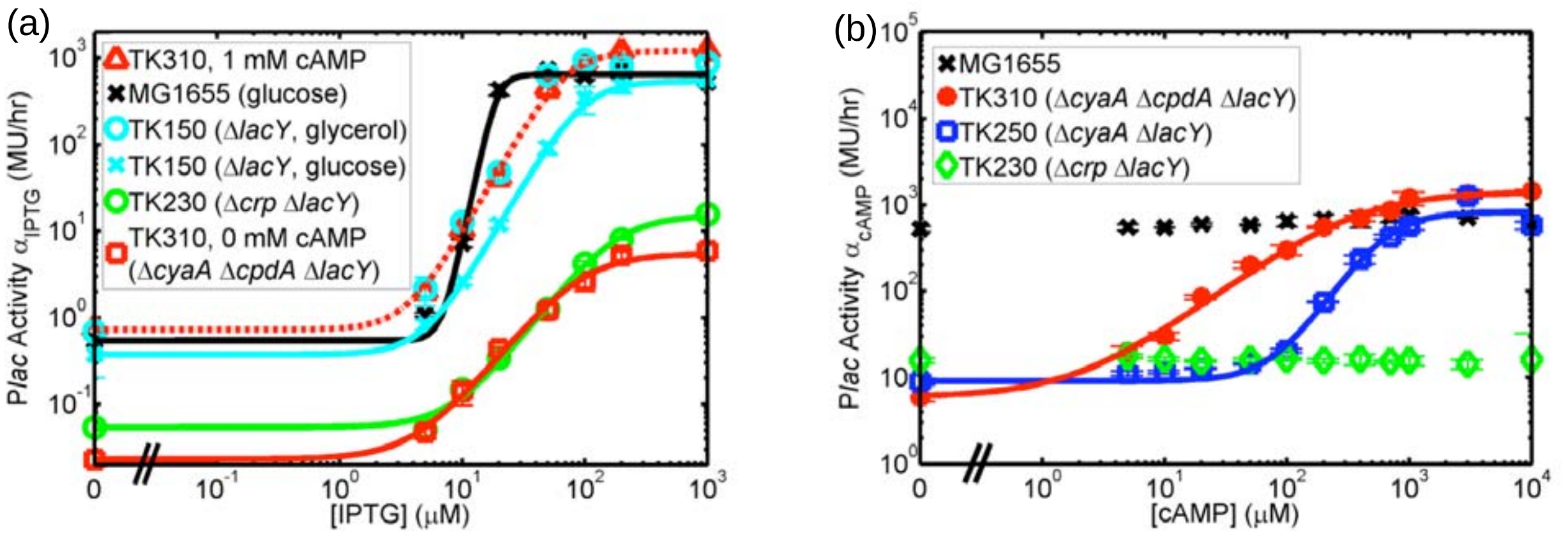

**Figure 1: Dependence of Plac activity on the inducers.** (a) The IPTG response for *E. coli* MG1655 and various Δ*lacY* mutants, grown in minimal M9 medium with varying amounts of IPTG and 0.5% glucose except for the cyan circles (0.5% glycerol). No cAMP was added to the medium except for the red triangles where 1mM cAMP was added. The lines are best fits to the Hill function (Eq. 1). (b) The cAMP response for *E. coli* MG1655 and various Δ*cyaA* mutants, grown in minimal M9 medium + 0.5% glucose, 1mM IPTG and varying amount of cAMP. The lines are best fits to the Hill function (Eq. 2).

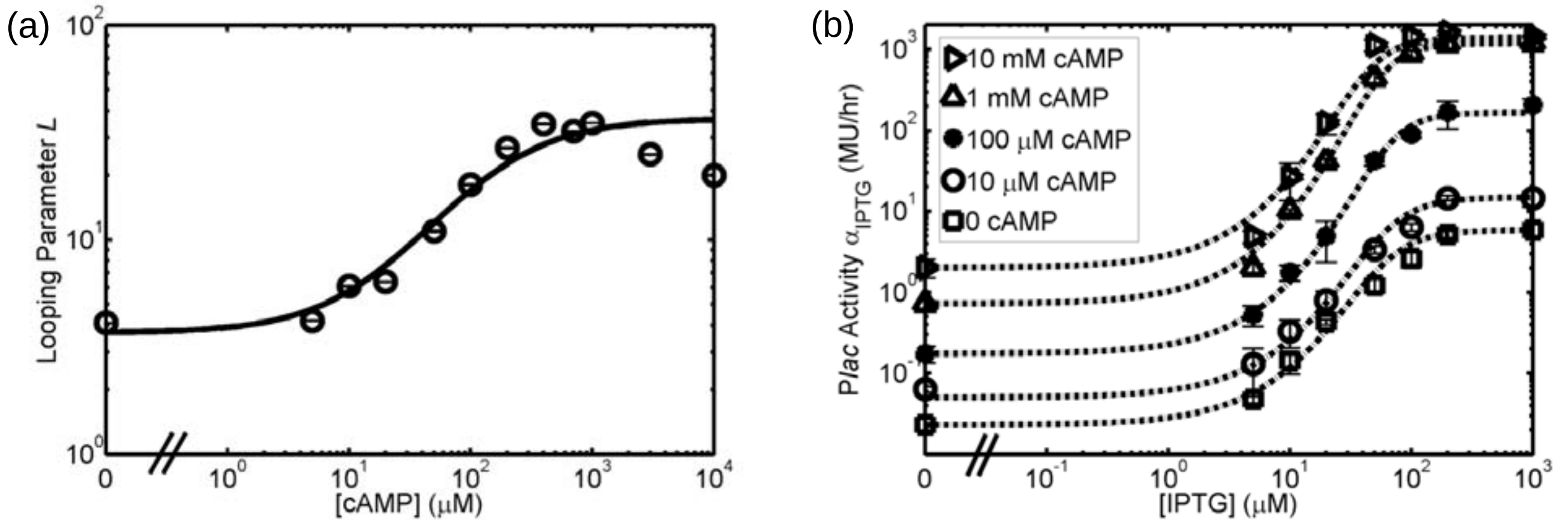

**Figure 2: Combinatorial control of Plac activity for strain TK310. (a)** The cAMP-dependent loop parameter $L$ is inferred from Eq. (4) by fixing the maximal fold-change $f_{\text{IPTG}}$ to be the ratio of $\alpha_0$ and the observed promoter activity in medium with no IPTG. $\alpha_0$ is generated by the bare cAMP response $\alpha_{\text{cAMP}}$ using the parameters in Table 2 row 2. The result obtained (circles) is fitted to Eq. (5) using a single fitting parameter $\Omega = 10.3 \pm 0.1$, with $C_{\text{cAMP}} = 320$ μM (Table 2 row 2) and $L_0 = 3.7$ (Table 3 row 4). **(b)** The IPTG responses obtained at different cAMP levels (symbols) are plotted together with predictions of the CRP-assisted DNA-looping model (Eq. 3) with *no* adjustable parameters. We used $R = 50$, $K_{\text{IPTG}} = 12.3$ μM (Table 3 row 3), with values of $L$ taken from (a) and $\alpha_{\text{cAMP}}$ for $\alpha_0$ (Eq. (2) with parameters values given in Table 2 row 2).

# TABLES

| Strain | $b_{IPTG}$ (MU/hr) | $f_{IPTG}$ | $C_{IPTG}$ (μM) | $m_{IPTG}$ |
|---|---|---|---|---|
| **MG1655** | 0.5 | 1200 ± 80 | 20 ± 4 | 4.5 ± 0.7 |
| **TK150 (Δ*lacY*)** | 0.4 | 1445 ± 185 | 100 ± 18 | 2.6 ± 0.2 |
| **TK230 (Δ*crp* Δ*lacY*)** | 0.05 | 255 ± 19 | 150 ± 15 | 2.0 ± 0.1 |
| **TK310 (Δ*cyaA* Δ*cpdA* Δ*lacY*), no cAMP** | 0.02 | 238 ± 26 | 90 ± 5 | 2.0 ± 0.2 |
| **TK310, 1 mM cAMP** | 0.7 | 1600 ± 180 | 70 ± 8 | 2.8 ± 0.1 |

**Table 1: Hill parameters for the IPTG response.** Parameters derived from fit of the IPTG dependence of the Plac activity to the Hill form (Eq. 1), for various strains of *E. coli* derived from MG1655, grown in medium with 0.5% glucose and varying amounts of IPTG. For the last row, 1mM cAMP was also added to the growth medium.

| Strain | $b_{cAMP}$ (MU/hr) | $f_{cAMP}$ | $C_{cAMP}$ (μM) | $m_{cAMP}$ |
|---|---|---|---|---|
| **TK250 (Δ*cyaA* Δ*lacY*)** | 9.1 | 91 ± 5 | 645 ± 43 | 2.1 ± 0.1 |
| **TK310(Δ*cyaA* Δ*cpdA* Δ*lacY*)** | 5.9 | 240 ± 13 | 320 ± 32 | 1.0 ± 0.1 |
| **prediction based on thermodynamics** | --- | > 20 | 10 – 1000 | 1 |

**Table 2: Hill parameters for the cAMP response.** The first two rows give parameters derived from fit of cAMP response of *cyaA* mutants to the Hill function (Eq. 2). The last row give best estimates of these parameters according to the thermodynamic model and the known biochemical parameters; see Supp Mat for details.

| Strain | $K_{IPTG}$ (μM) | $L_0$ |
|---|---|---|
| **TK230 (Δ*crp* Δ*lacY*)** | 15.7 ± 1.5 | 4.1 |
| **TK310 (Δ*cyaA* Δ*cpdA* Δ*lacY*)** | 12.3 ± 1.4 | 3.7 |
| **TK250 (Δ*cyaA* Δ*lacY*)** | 17.8 ± 1.7 | 5.0 |

**Table 3: Parameters for model with DNA Looping.** Parameters derived from the fit of IPTG response of given strains to the thermodynamic model with DNA looping (Eq. 3).

# SUPPLEMENTARY TABLES AND FIGURES

| Strain | Genotype | Derived From | Comments |
|---|---|---|---|
| TK110 | Δ*cpdA* | MG1655 | deletion from 3174100 to 3174793 |
| TK120 | Δ*crp* | MG1655 | deletion from 3483856 to 3484369 |
| TK130 | Δ*cyaA* | MG1655 | deletion from 3988871 to 3991173 |
| TK140 | Δ*lacI* | MG1655 | deletion from 365748 to 366642 |
| TK150 | Δ*lacY* | MG1655 | deletion from 361259 to 362274 |
| TK200 | Δ*lacI* Δ*lacY* | TK140 | |
| TK210 | Δ*cpdA* Δ*lacY* | TK110, TK150 | |
| TK220 | Δ*crp* Δ*lacI* | TK110, TK140 | |
| TK230 | Δ*crp* Δ*lacY* | TK120, TK150 | |
| TK240 | Δ*cyaA* Δ*lacI* | TK130, TK140 | |
| TK250 | Δ*cyaA* Δ*lacY* | TK130, TK150 | |
| TK310 | Δ*cyaA* Δ*cpdA* Δ*lacY* | TK110, TK130, TK150 | |
| TK320 | Δ*cyaA* Δ*lacI* Δ*lacY* | TK130, TK200 | |
| TK330 | Δ*crp* Δ*lacI* Δ*lacY* | TK120, TK200 | |
| BW25113 | $lacI^{q}$ $rrnB_{T14}$ $\Delta lacZ_{WJ16}$ *hsdR514* $\Delta araBAD_{AH33}$ $\Delta rhaBAD_{LD78}$ | K12 BD792 | Datsenko and Wanner, 2000 |
| TK400 | Δ*cyaA* | BW25113 | deletion from 3988871 to 3991173 |
| TK500 | Δ*cyaA* Φ(P*crp*-*lacZ*) | TK400 | P*crp* amplified from 3483424 to 3483721 |
| TK600 | Δ*cyaA* Φ(P*lac*Δ*crp*-*lacZ*) | TK400 | CRP operator at -61.5 altered from TGTGAGTTAGCTCACT to CAGACGTTAGCTCACT |

**Table S1: List of strains and plasmids used in this study.**

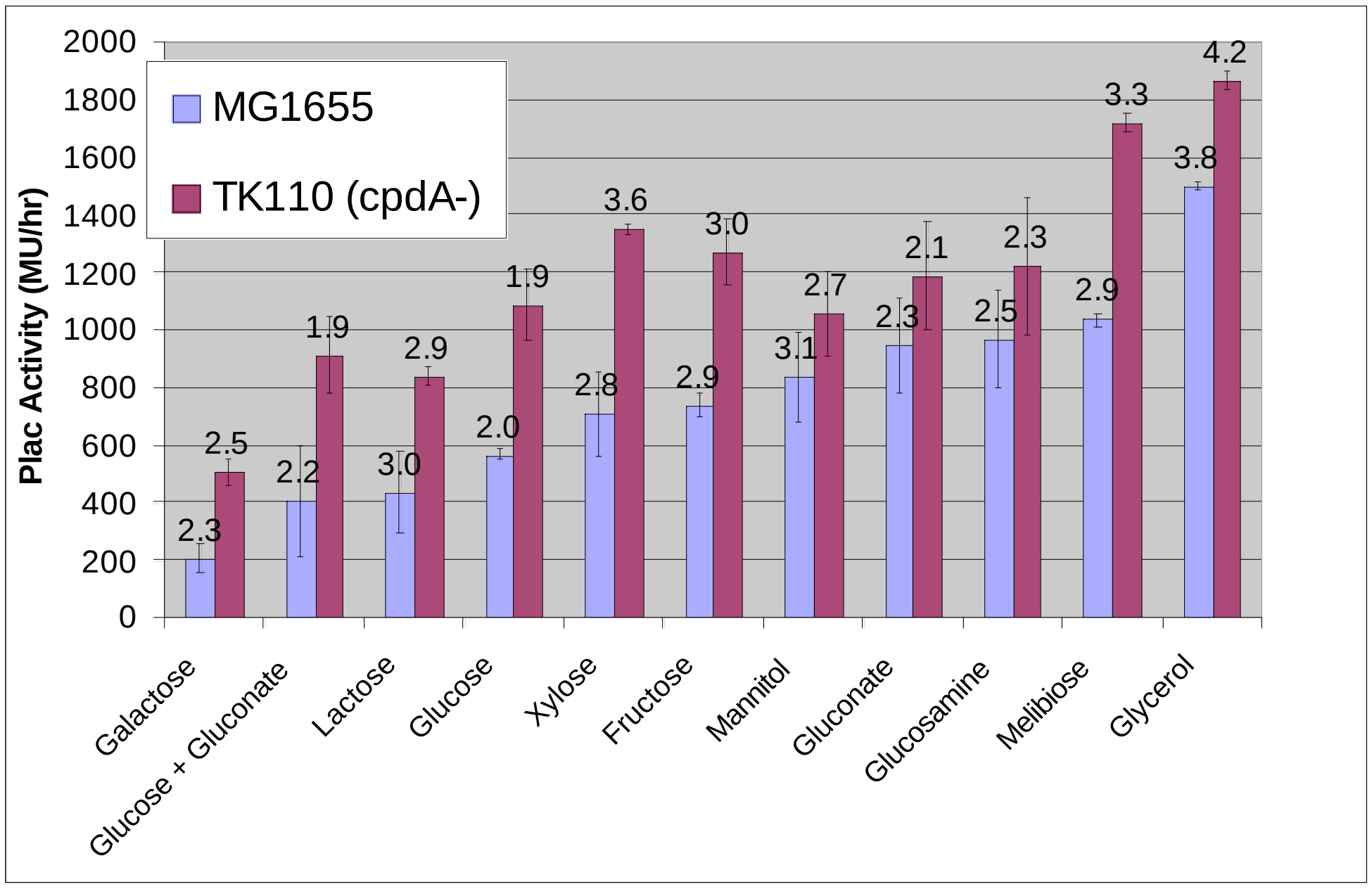

**Figure S1: Plac activity of *E. coli* MG1655 cells and *ΔcyaA* mutants grown with various sugar sources.** The cells are grown in minimal M9 medium with 1mM IPTG and with 0.5% of the indicated sugars as the sole carbon source. The number above each bar is the observed doubling times (in hr) for that strain and medium.

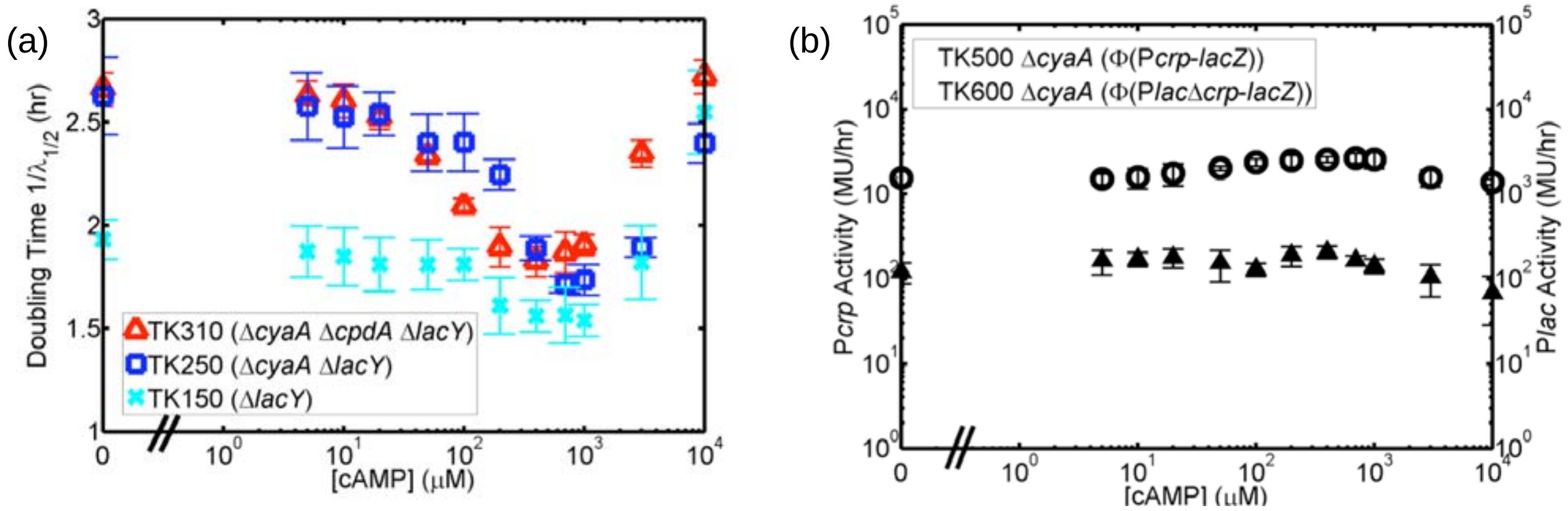

**Figure S2: Indirect effect of cAMP. (a)** The doubling time $1/\lambda_{\frac{1}{2}}$ of various mutant strains of *E. coli* MG1655 grown in M9 minimal medium with 0.5% glucose and varying amount of cAMP. For the *cyaA*⁻ strains (blue and red symbols), the growth rate varied by as much as 40%, with the most rapid growth in medium with 1mM cAMP. **(b)** The activity of the *crp* promoter was determined by the β-galactosidase assay for strain TK500 (BW25113 cells bearing *cyaA* deletion and chromosomal insertion of the *crp* promoter transcriptionally fused to *lacZ*.). The activity was found to vary by no more than 2-fold in medium containing 0 to 10 mM cAMP, with maximal activity at ~ 1mM cAMP. The *indirect* effect of cAMP-CRP on the *lac* promoter was determined by the β-galactosidase assay for strain TK600 (BW25113 cells bearing *cyaA* deletion and chromosomal insertion of a mutated *lac* promoter:*lacZ* construct, with the promoter containing a mutated CRP operator site; see Table S1). The activity of this promoter was nearly unchanged in media containing 0 to 1mM cAMP, and decreasing only by 2-fold in medium with 1 to 10 mM cAMP. (1mM IPTG was added to the medium to alleviate repression by LacR.) The results indicate that the cAMP dependence observed for wildtype *lac* promoter (Fig. 1a in the main text) are primarily due to cAMP-CRP exerted through the CRP operator site.

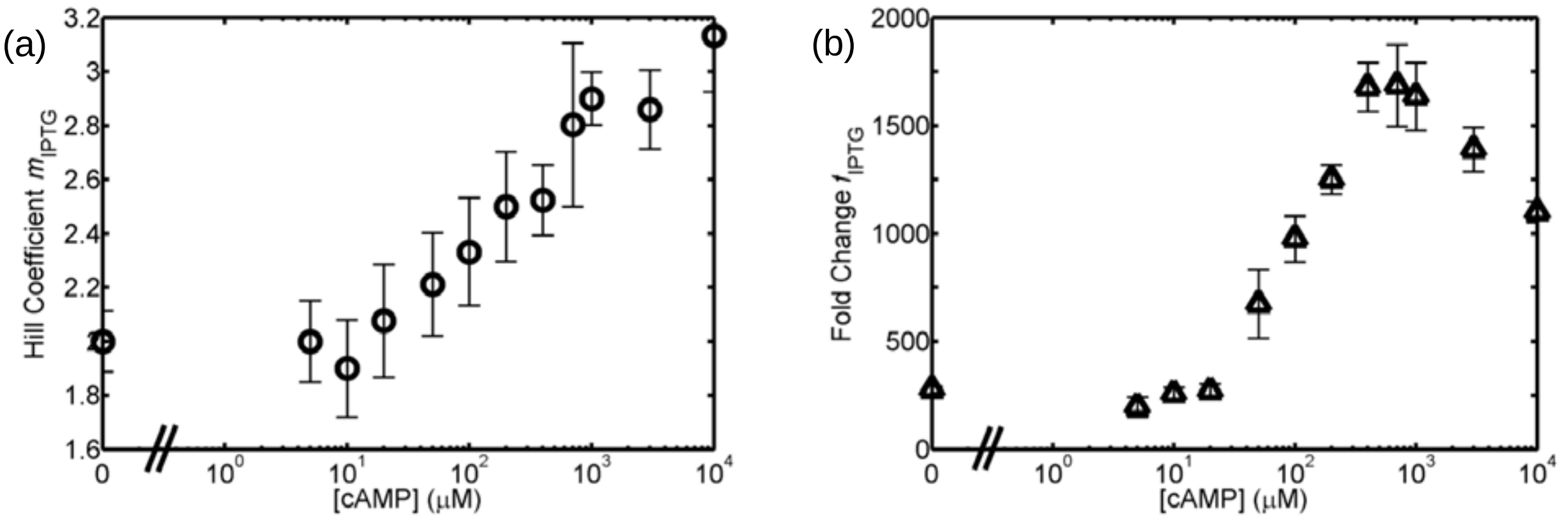

**Figure S3: cAMP dependence of IPTG response.** Values of the Hill parameters $m_{IPTG}$ (a) and $f_{IPTG}$ (b) obtained from fitting the IPTG response of TK310 cells grown in media with the indicated cAMP concentrations

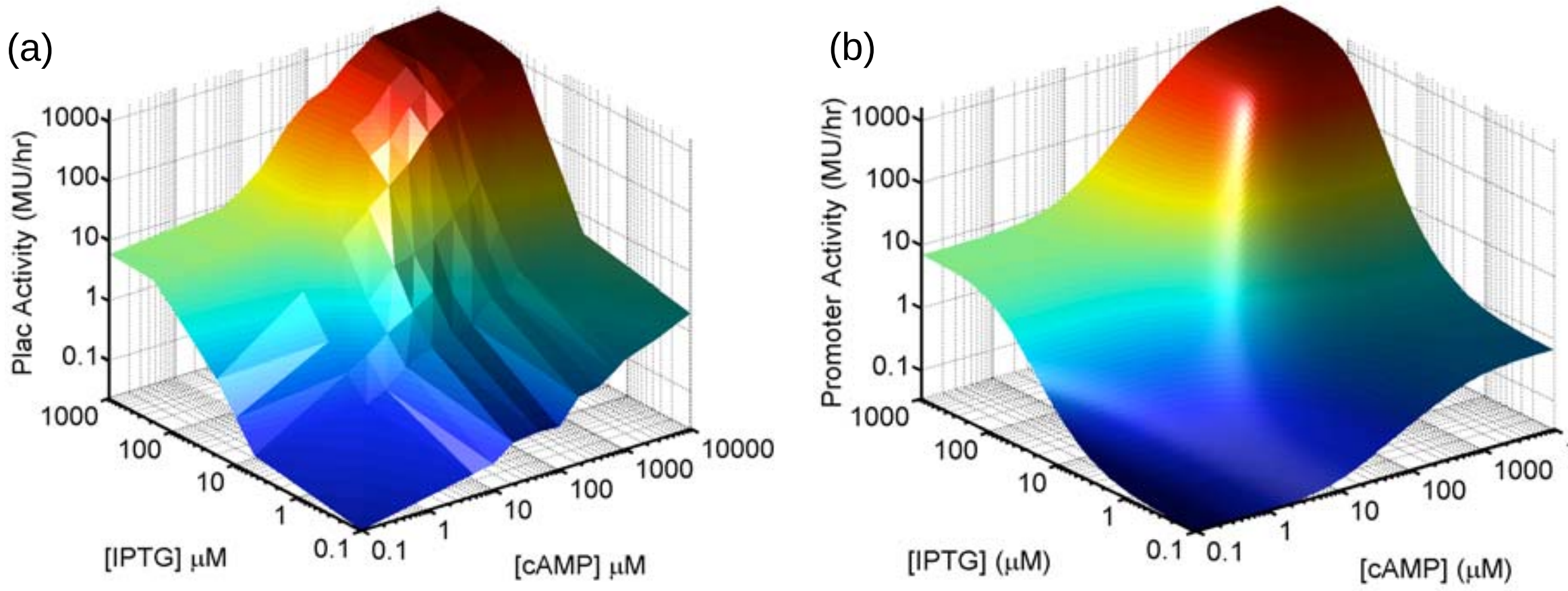

**Figure S4: Combinatorial control of Plac activity for strain TK310. (a)** 3D plot of Plac activity for strain TK310 (Δ*lacY* Δ*cyaA* Δ*cpdA*) grown in M9 medium with varying amounts of cAMP and IPTG. **(b)** Predicted form of Plac activity according to the extended thermodynamic model. The function is generated using only 5 parameters: $K_{\mathrm{IPTG}}$ and $L_0$ (Table 3 row 2) for the bare IPTG response, $C_{\mathrm{cAMP}}$ and $f_{\mathrm{cAMP}}$ (Table 2 row 2) for the bare cAMP response, and $\Omega$ = 10.3 for the co-dependence.

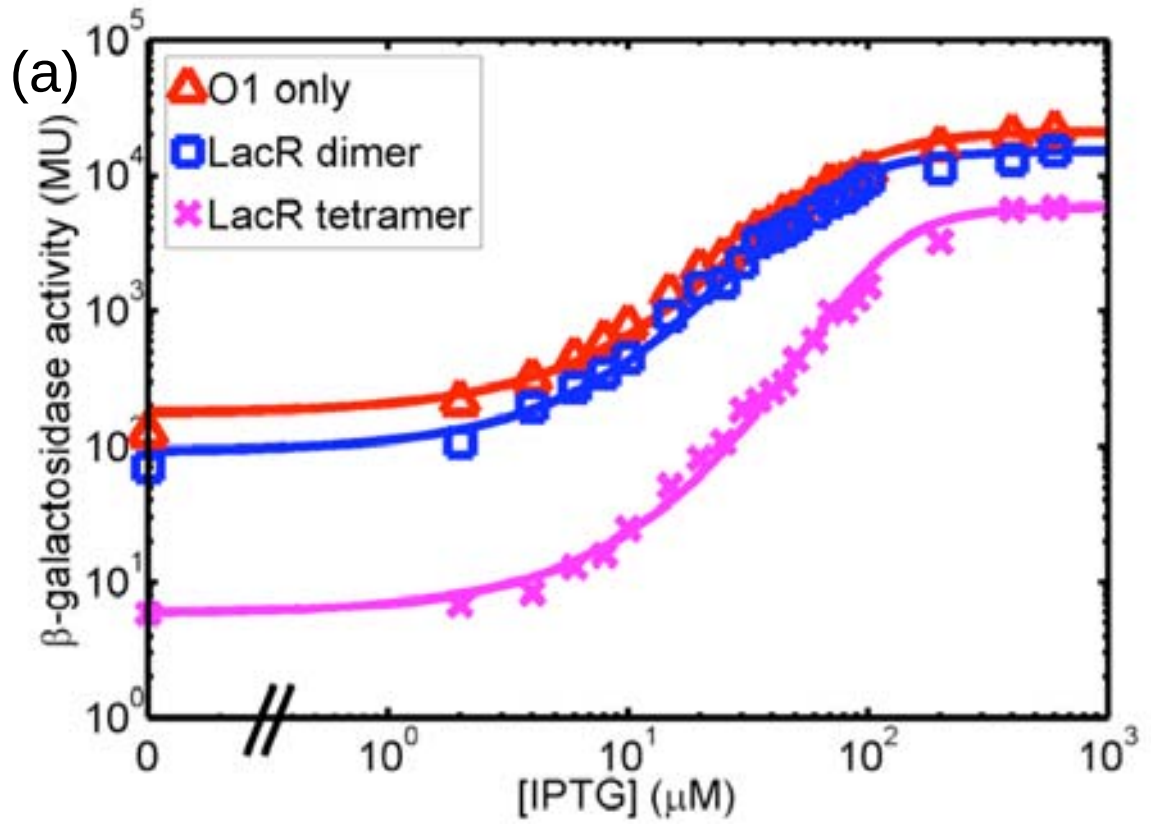

**Figure S5: IPTG response in the absence of DNA looping.** Data from Oehler et al, 2006 (62) with overexpressed LacR. Strains include wildtype promoter and repressor (LacR tetramer; magenta crosses), wildtype promoter with dimeric repressor (LacR dimer; blue squares), and wildtype repressor with promoter including O1 binding site only (O1 only; red triangles); (a) fit of the nonlooping mutants, red triangles and blue squares, to Eq. (3) with $L_0 = 0$, yielding $R \sim 200$ and $K_{\mathrm{IPTG}} \sim 6$ μM. The wildtype fit to the looping model (with R = 200 and $L_0$ fixed by the fold-change); (b) fit of the same data to the Hill form. O1 only and LacR dimer strains yield Hill coefficient $m_{\mathrm{IPTG}} = 1.5$; wildtype strain yields $m_{\mathrm{IPTG}} = 2.5$.

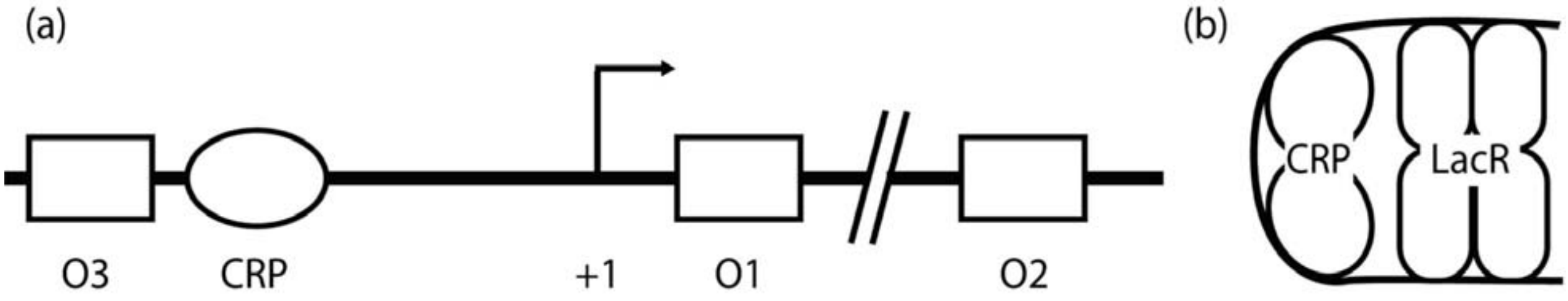

**Figure S6: Schematics of the *lac* promoter. (a)** The main Lac operator O1 is located at position +11 from the transcriptional start. The auxiliary Lac operator O2 is located at +411 and O3 at -81. The CRP operator, at position -62.5, is located in between O1 and O3. **(b)** Illustration of the hypothesized interaction between CRP and LacR: The binding of CRP to its operator site bends the DNA in a way to facilitate the looping between O1 and O3.

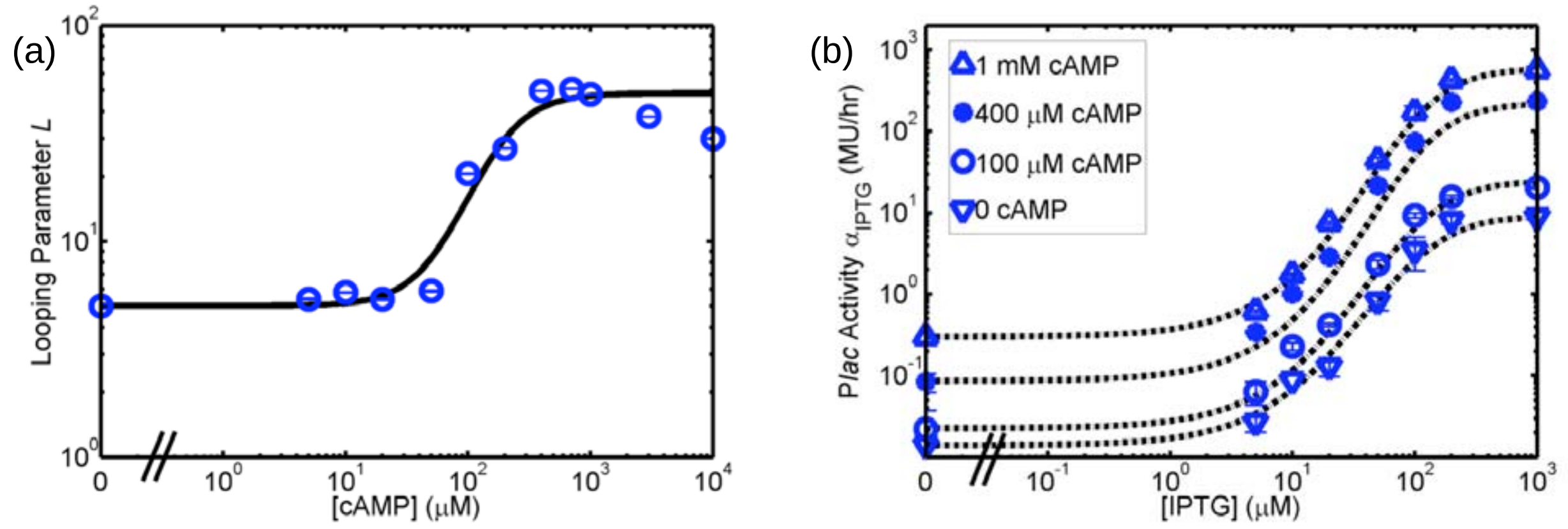

**Figure S7: Combinatorial control of Plac activity for strain TK250. (a)** The cAMP-dependent loop parameter $L$ is inferred from Eq. (4) by fixing the maximal fold-change $f_{\text{IPTG}}$ to be the ratio of $\alpha_0$ and the observed promoter activity in medium with no IPTG. $\alpha_0$ is generated by the bare cAMP response $\alpha_{\text{cAMP}}$ using the parameters in Table 2 row 1. The result obtained (circles) is fitted to Eq. (S26) using a single fitting parameter Ω, determined to be 9.7, with $C_{\text{cAMP}}$ = 645 μM (Table 2 row 1) and $L_0$ = 5.0 (Table 3 row 3). **(b)** The IPTG responses obtained at different cAMP levels (symbols) are plotted together with predictions of the DNA-looping model (Eq. S15) with no adjustable parameters. We used $R$ = 50, $K_{\text{IPTG}}$ = 17.8 μM (Table 3 row 3), with values of $L$ taken from (a) and $\alpha_{\text{cAMP}}$ (Eq. 2 with parameters in Table 2 row 1) for $\alpha_0$; see Supp Mat.

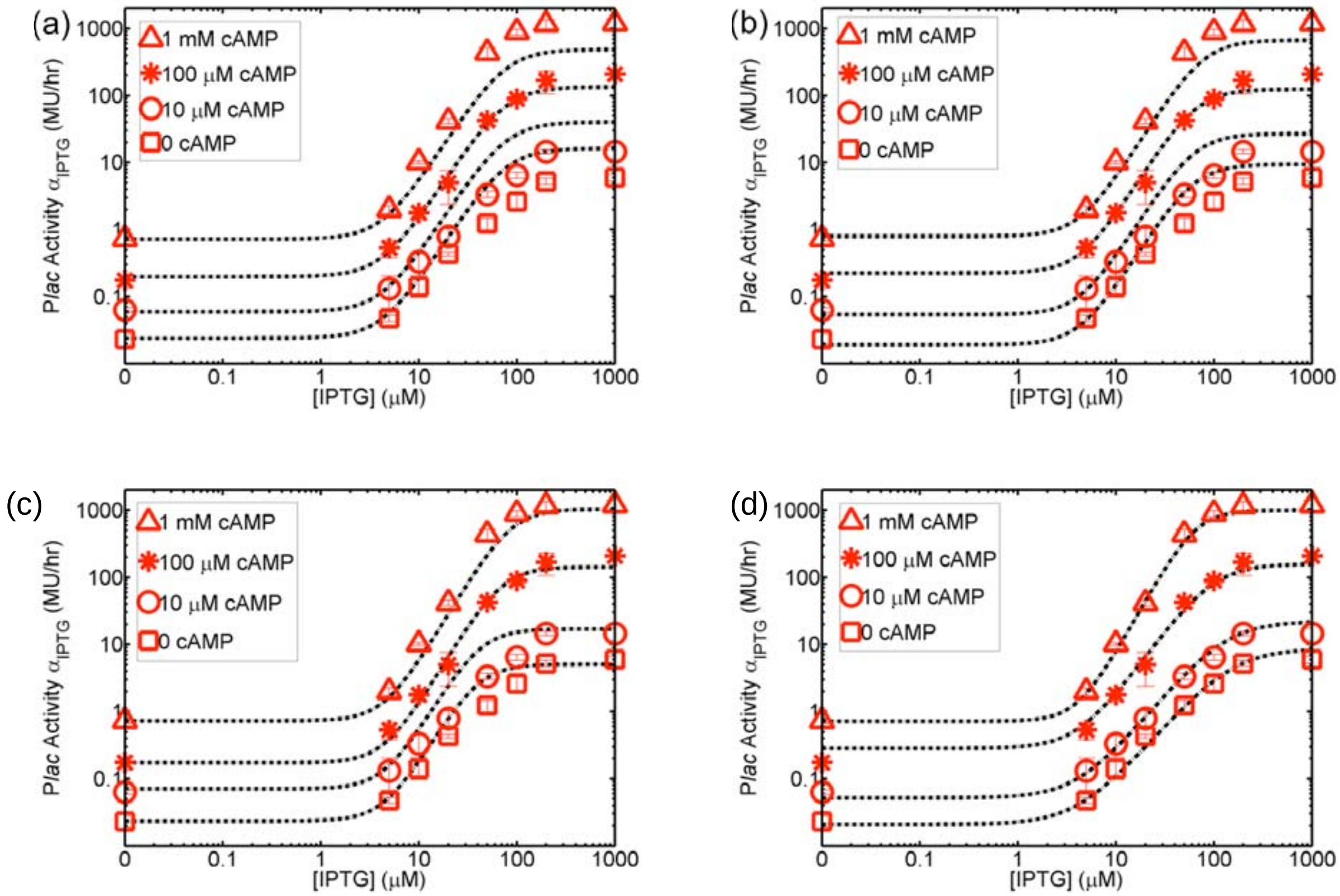

**Figure S8: Comparison to alternative models.** (a) A naïve model assuming no interaction between CRP and LacR. (b) A model assuming cAMP dependent fold change with $C'_{cAMP} = C_{cAMP}$, $m'_{cAMP} = m_{cAMP}$ (see Eq. **Error! Reference source not found.**).
(c) Same as (b), but with freely varying cAMP related parameters. (d) Same as (b), but assuming a LacR-IPTG Hill coefficient, $m_{IPTG}$, governed by a Hill function (Eq. **Error! Reference source not found.**).

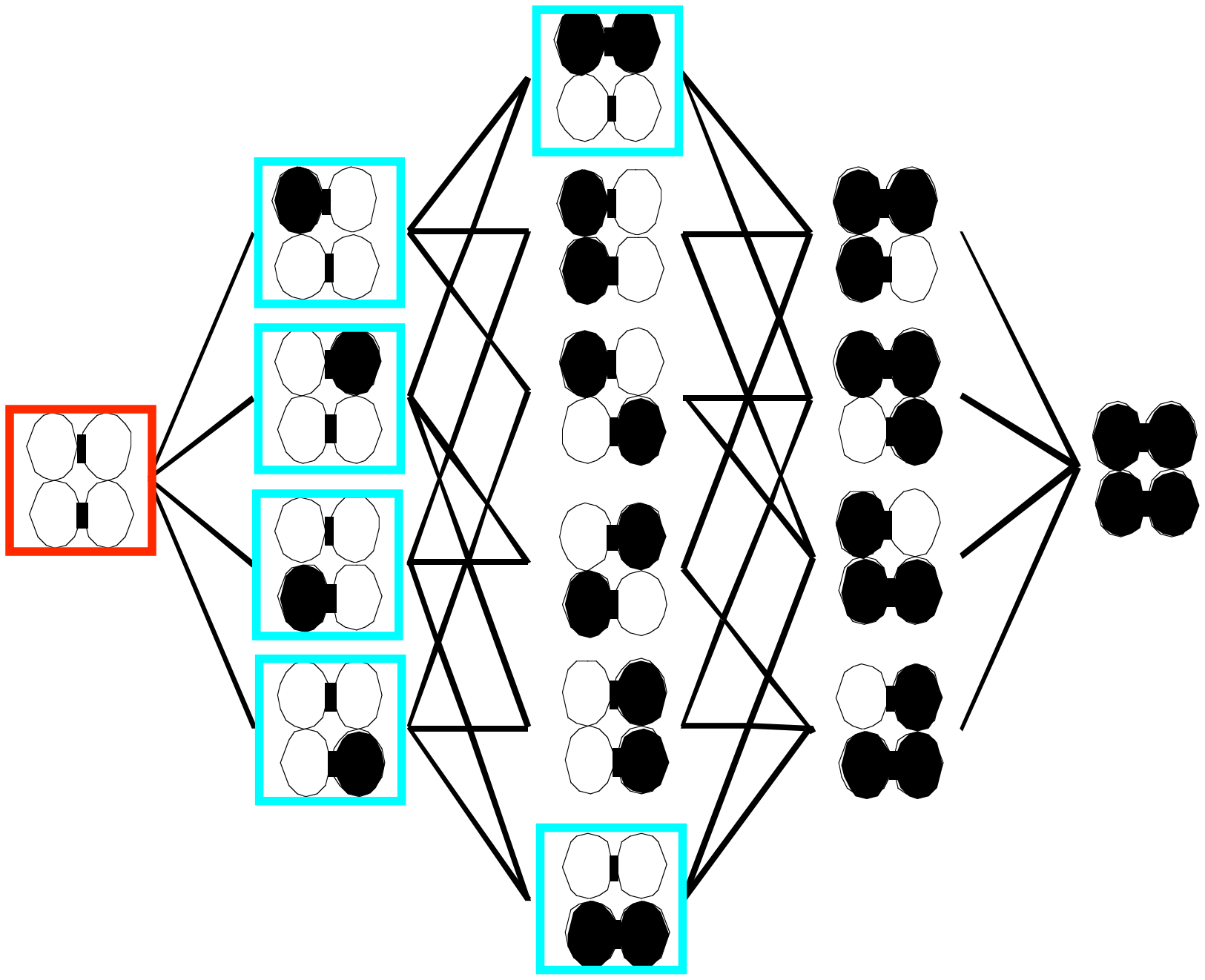

**Figure S9: Different states of LacR-IPTG binding.** The ellipses represent the LacR monomers. Two monomers form a dimeric unit (joined by the thick solid line). Two dimers form a tetramer. The open ellipses indicate LacR monomers not bound to IPTG; filled ellipses indicate LacR monomers bound to IPTG. Each line indicates an allowed direct transition between two states. From left to right, the different columns represent subspecies denoted as LacR-0, LacR-1, LacR-2, LacR-3, and LacR-4, respectively in Supp Mat. The active species LacR** and LacR* are indicated by the red and cyan boxes respectively.

# SUPPLIMENTARY MATERIALS

# EXPERIMENTAL METHODS

**a. Plasmids and Strains.** *E. coli* K-12 MG1655 and derived mutants were used in all experiments reported in the main text. Chromosomal deletions were performed using the method of Datsenko & Wanner (1). For each deletion the kanamycin resistance gene *kan* was amplified from the template plasmid pKD4 using primer sites P0 and P2 including 45 bp homology extensions. The PCR products were then transformed into electrocompetent MG1655 cells, and *kan* was eliminated using plasmid pCP20 bearing FLP recombinase (1). The resulting strains TK110, TK120, TK130, TK140, and TK150 are listed in Table S1 together with the start and end coordinates of the respective deletions. All mutations were verified using PCR. Multiple deletion strains were constructed by serially transferring and eliminating the TK100 series of *kan* insertion mutants into the target strain using P1 transduction in the order indicated in column 3 of Table S1, with the exception of strain TK200 carrying Δ*lacI* Δ*lacY* double mutation. The close proximity of *lacI* to *lacY* necessitated the *de novo* synthesis of both mutations in the same host strain.

Measurements of the variation of CRP levels were performed on strain TK500. This strain was derived from strain TK400, which was derived in turn from *E. coli* BW25113 (Datsenko & Wanner, 2000), with *cyaA* deletion performed as explained above. The *crp* promoter P*crp* was amplified from MG1655 genomic DNA from start coordinate 3483424 to end coordinate 3483721 and inserted into the EcoRI and BamHI sites of PRS551 (2). The resulting P*crp-lacZ* fusion was inserted into the chromosome of TK400 using the method of Simons et al (1987). Measurements of the effect of the CRP operator on the *lac* promoter were performed using strain TK600 which is also derived from strain TK400. The *lac* promoter was amplified from start coordinate 365438 to end coordinate 365669 and cloned into the EcoRI and BamHI sites of pRS551. The CRP operator site at position -61.5 was altered from TGTGAGTTAGCTCACT to CAGACGTTAGCTCACT using site-directed mutagenesis (Stratagene) and the resulting construct was inserted into the chromosome of strain TK400 using the method of Simons et al (2).

**b. β-galactosidase Assay.** Assays of β-galactosidase activity were performed according to Miller (3) and Griffith (4) with changes as follows. Overnight cultures of bacterial strains were grown in 3ml M9 minimal medium (M9, 1 mM $MgSO_4$, 0.1 mM $CaCl_2$) containing 0.5% glucose as the sole carbon source and the standard concentrations of any necessary antibiotics in a 37°C water bath until saturation. This culture was diluted[1] into 24 well plates (Costar) containing 1ml M9 medium in each well with glucose, antibiotics, and varying concentrations of 3',5'-cyclic adenosine monophosphate (cAMP) (0-10 mM) and isopropyl β-D thiogalactopyranoside (IPTG) (0-1 mM). The plates were then grown with vigorous shaking at 200 rpm in a humidity-controlled incubator, with $OD_{600}$ measurements taken every two hours in a Tecan Genios Pro plate reader. The cell-doubling rate ($\lambda_{\frac{1}{2}}$) for each sample was calculated as the slope of $\log_2(OD_{600})$ vs. time plot via linear regression analysis.

When $OD_{600}$ of a sample reaches 0.2-0.4, 0.5 ml of the sample was transferred to a 2-ml 96 well polypropylene block containing 0.5 ml Z-buffer (3), 20 µl 0.1% SDS, and 40 µl chloroform. All samples were thoroughly disrupted by repeated agitation with a multichannel pipettor. 200 µl of each sample was transferred to a flat-bottom transparent 96 well plate (Costar). 40 µl of Z-buffer containing 4 mg/ml *o*-nitrophenyl-β-D-galactopyranoside (ONPG) was added to each well and $OD_{420}$ measurements were performed in a Tecan Genios Pro plate reader, at 1-minute intervals for the first 10 minutes, and at increasing intervals (1 hour, 5 hour, etc.) thereafter[2] until sufficient color had developed in all samples for measurement, and the measured $OD_{420}$ showed an extended regime of linear dependence on time (data not shown). Samples were maintained at 25°C throughout incubation. All assays were performed in triplicate or more.

**c. Promoter activity.** In the regime where the $OD_{420}$ readings were sufficiently large to ensure a linear dependence on time of measurements but below 1.0 (above which the $OD_{420}$ readings saturate), the slope ($s$) of $OD_{420}$ versus time (in minutes) was calculated via linear regression. β-galactosidase activity ($A$) was expressed in Miller Units (MU) according to the formula $A = (1000 \cdot s) / (0.5 \cdot \mathrm{OD}_{600})$. To verify the reliability of the deduced activity for the weakly expressed promoters, we performed a serial dilution experiment, using MG1655 grown in M9 medium with 0.5% glucose and 1 mM IPTG, which yielded a β-galactosidase activity of ~ 1000 MU. We mixed this strain with strain BW25113, deleted of the *lac* operon, in varying proportions (up to $10^6$-fold dilution), and measured the apparent β-galactosidase activity of the mixture using the procedure specified above. We found a nearly perfect inverse dependence of the observed β-galactosidase activity with the fold-dilution applied down to ~ 0.03 MU (data not shown). As the lowest β-galactosidase activity encountered in the experiments was ~0.1 MU, our data lay completely in the responsive regime of the measurements.

---

[1] In M9 medium, the growth rate of Δ*cyaA* strains increases from 0.4 doubling/hour at 0 cAMP to a maximum of 0.6 doubling/hour at ~1 mM cAMP; see Fig. S2. As such, the initial inoculation of samples was performed at different dilutions (~1000x dilution for samples in 1 mM cAMP and ~ 100x dilution for samples at lower cAMP concentrations) such that all samples would reach $OD_{600}$ = 0.2-0.4 in approximately 12-15 hours. We verified that samples assayed with various initial dilutions in the given range gave no measurable systematic variations in results (data not shown).

[2] For the lowest expression encountered in the experiments performed (e.g., the *crp* null mutants in Fig. 1a), this took up to 50 hours.

In the text, we report the "promoter activity" (α) as

$$\alpha = A \cdot \lambda_{\frac{1}{2}} \tag{S1}$$

where $\lambda_{\frac{1}{2}}$ is the cell-doubling rate defined above (in unit of 1/hr). This measure of the promoter activity is motivated by the fact that the enzyme β-galactosidase is very stable (5) so that in the balanced exponential growth phase, its "turnover" is governed by dilution due to cell growth; see discussion below. Note that with the definition (S1), we take into account some straightforward growth-dependent effects on β-galactosidase activity as previously reported (6, 7). There are of course other residual effects, e.g., variation in the cellular levels of RNA polymerase, ribosomes, etc at the different growth rates. However, these effects are apparently quite limited in magnitude and will be neglected in our analysis: As seen in Fig. S2b (triangles), the growth-rate adjusted activity of β-galactosidase expressed by the (constitutive) P*lacΔcrp* promoter in *cyaA*- cells changed very little in growth medium with various cAMP concentrations, even though the growth rate itself changed appreciably over this range of cAMP concentrations (Fig. S2a).

# THEORETICAL MODELING

**a. Activation of the Regulatory Proteins.** The concentrations of the active regulators are controlled by the total concentrations of the regulators, the intracellular inducer concentrations, and the biochemistry of inducer-regulator interaction. The interaction of CRP with cAMP is quite straightforward (8-11) and will be discussed first. The active component is the CRP dimer associated with *one* cAMP molecule (10-12). For simplicity, we will assume that all the CRP monomers associate in the dimer form, as justified by the very small dimer dissociation constant (0.1 – 1 nM; (13)) The concentration of the activated CRP dimer (denoted by $[CRP^*]$) is then given by

$$\left[CRP^*\right] \equiv [cAMP - CRP] = [CRP] \cdot \frac{[cAMP]_{in}^*}{[cAMP]_{in}^* + K_{cAMP}}, \qquad \text{(S2)}$$

where [CRP] denotes the *total* CRP dimer concentration, $K_{cAMP} = 10\ \mu$M is the relevant dissociation constant (10), and $[cAMP]_{in}^*$ is the intracellular concentration of cAMP. Intracellular cAMP is believed to be rapidly exported by PMF pumps (14, 15). For *cyaA* strains which cannot synthesize cAMP endogenously, the high export/import ratio leads to a much smaller intracellular cAMP concentration compared to the extracellular concentration. It is this intracellular cAMP concentration which dictates the degree of CRP activation (assuming that cAMP-CRP equilibration is much faster than cAMP transport.) The details of cAMP transport are not understood; even the cAMP pump has not been identified despite many years of study. Here, we appeal to the results of Epstein et al (16) and assume that the intracellular cAMP concentration is linearly related to the extracellular concentration across the range of concentrations used, i.e., $[cAMP]_{in}^* = \gamma_{cAMP} \cdot [cAMP]$, where [cAMP] is the cAMP level of the medium. Since $\gamma_{cAMP} \approx 10^{-3}$ according to Ref. (16), we have $[cAMP]_{in}^* = \gamma_{cAMP} \cdot [cAMP] \ll K_{cAMP}$ for most of the range of cAMP used in our experiment (0 – 10 mM). Hence Eq. (S2) is simplified to

$$\left[CRP^*\right] \approx [CRP] \cdot [cAMP] \cdot \gamma_{cAMP} / K_{cAMP}. \qquad \text{(S3)}$$

Obtaining the active concentration of the Lac repressor is more involved. The Lac repressor is a dimer of dimers (17). Co-crystal structure of LacR and operator DNA indicates that the dimer binds specifically to the operator sequence as a unit in the absence of IPTG, while the IPTG-bound dimers do not have the required structure to bind specifically to operator DNA. Additionally, since the two dimeric units of the LacR tetramer interact very weakly (18), we do not expect the IPTG binding by one dimeric unit to affect the specific DNA binding of the other dimeric unit. Finally, we do not expect a dimeric unit to be able to bind specifically to operator sites if either one of its monomeric subunit is bound to IPTG. This assumption is based on the work of Winter, Berg and von Hippel (19), who showed that sequence specific binding would only occur if the energy gained from sequence-specific binding more than compensates the loss of electrostatic energy the repressor molecule could gain in a conformation that does not allow specific DNA binding. When one of the monomers bind to IPTG, that monomer could not contribute to specific DNA binding, and the other monomer would not be able to contribute enough to compensate given what we know about the energetics of LacR-DNA binding (19). The last feature described above, that both monomers of a dimeric unit need to be free of IPTG binding in order for the dimeric unit to bind specifically to operator DNA, is a crucial source of cooperativity in our model. While there is no direct biochemical proof of this feature, we mention that a MWC model incorporating this feature of LacR-DNA binding (Kuhlman and Hwa, unpublished) is able to reproduce the effective Hill coefficient of 1.4 ~ 1.6 which describes the IPTG dependence of specific operator binding, as observed in the experiments of O'Gorman et al (20). It then follows that there are two species of "active repressors" in our model: (i) Only one of the two dimeric units is not bound at all by IPTG (denoted by LacR*), and (ii) neither of the dimeric units is bound by IPTG (denoted by LacR**).

To find the concentrations of these two active species (denoted by $[LacR^{*}]$ and $[LacR^{**}]$ respectively) in terms of the total LacR tetramer concentration [LacR] and the IPTG concentration $[IPTG]$, we need to keep track of all possible molecular species of LacR-IPTG complex. As illustrated in Fig. S9, there is one species with no IPTG (denoted by LacR-0), 4 sub-species with one IPTG bound (LacR-1), 2 sub-species with two IPTG bound to a single dimer (LacR-2a), 4 sub-species with two IPTG bound to two different dimers (LacR-2b), 4 subspecies with 3 IPTG bound (LacR-3) and one subspecies with 4 IPTG bound (LacR-4). Since the IPTG-LacR binding is non-cooperative (20), the equilibrium between any two subspecies linked by a line in Fig. S9 is described by the *same* dissociation constant $K_{IPTG}$. Taking into account of the indistinguishability of the subspecies, we arrive at the following equilibrium relations among the species of LacR-IPTG complex:

$$\frac{[LacR-1]}{[LacR-0]}=\frac{4[IPTG]}{K_{IPTG}};\ \frac{[LacR-2a]}{[LacR-0]}=2\cdot\left(\frac{[IPTG]}{K_{IPTG}}\right)^2;\ \frac{[LacR-2b]}{[LacR-0]}=4\cdot\left(\frac{[IPTG]}{K_{IPTG}}\right)^2;$$
$$\frac{[LacR-3]}{[LacR-0]}=4\cdot\left(\frac{[IPTG]}{K_{IPTG}}\right)^3;\ \frac{[LacR-4]}{[LacR-0]}=\left(\frac{[IPTG]}{K_{IPTG}}\right)^4$$

Since the total LacR concentration is the sum of the concentrations of the individual species, then we have

$$\left[LacR^{**}\right]\equiv[LacR-0]=[LacR]\Big/\left(1+[IPTG]/K_{IPTG}\right)^4 \quad \text{(S4)}$$

and

$$\left[LacR^{*}\right]\equiv[LacR-1]+[LacR-2a]=[LacR]\cdot\frac{4\left([IPTG]/K_{IPTG}\right)+2\left([IPTG]/K_{IPTG}\right)^2}{\left(1+[IPTG]/K_{IPTG}\right)^4}. \quad \text{(S5)}$$

**b. Thermodynamic Modeling.** The transcriptional activity of a given promoter controlled by the binding and interaction of transcription factors (TF) can be modeled using a generalized Shea-Ackers like thermodynamic model (21) as described in Buchler et al (22) and Bintu et al (23, 24). Briefly, we assume that the rate of protein expression, *G*, is proportional to the equilibrium probability *P* of the RNA polymerase (RNAP) binding to its DNA target, the promoter, given the cellular concentrations of all of the relevant transcript factors. The probability *P* can be written as

$$P=\frac{Z_{ON}}{Z_{ON}+Z_{OFF}}$$

where $Z_{ON}$ and $Z_{OFF}$ are the partition sums of the Boltzmann weights over all states of transcription factor binding for the promoter bound and not bound, respectively, by the RNA polymerase. Given the knowledge of the cellular concentrations of the TFs, the partition function of the TFs themselves ($Z_{OFF}$) is simply

$$Z_{OFF}=\sum_{all\ \sigma_i=\{0.1\}}\prod_{(i,j)}\left(\frac{[TF_{s(i)}]}{K_i}\right)^{\sigma_i}\cdot\left(\frac{[TF_{s(j)}]}{K_j}\right)^{\sigma_j}\cdot\omega_{i,j}^{\sigma_i\sigma_j} \quad \text{(S6)}$$

where $K_i$ refers to the dissociation constant of operator site $i$ with its TF [of type *s* as indicated by subscript $s(i)$], and $\omega_{i,j}$ is the cooperativity factor between the two TFs bound to operator sites *i* and *j*. It is important to note that $K_i$ is the effective *in vivo* binding constant including the effect of binding of its TF to all other possible binding sites on the chromosome, and is typically not the value of *in vitro* measurement (25). The values of the $\omega_{i,j}$ depend on the spacing between sites *i* and *j*. If two sites are overlapping, $\omega_{i,j} = 0$, reflecting the fact that they cannot be simultaneously occupied. If two sites are positioned such that the two bound TFs can not contact each other, then $\omega_{i,j} = 1$, indicating the lack of interactions between the bound TFs. For two TFs that can bind cooperatively and their binding sites are positioned such that the bound TFs do interact, then the value of $\omega_{i,j}$ is the cooperativity factor of that interaction, and is typically of the order 10~100 (22). The expression for $Z_{ON}$ is similar to (S6), but with one of the sites being the promoter, and one of the [$TF_s$] being [RNAP].

As an example, consider a promoter with a single binding site for an activator protein *A*, which binds to its site with dissociation constant $K_A$ and interacts with RNAP via a cooperativity factor $\omega > 1$. The equilibrium promoter occupation probability *P* can then be written as

$$P = \frac{[RNAP]/K_{RNAP} \cdot (1 + \omega \cdot [A]/K_A)}{1 + [A]/K_A + [RNAP]/K_{RNAP} \cdot (1 + \omega \cdot [A]/K_A)} \simeq \frac{[RNAP]}{K_{RNAP}} \cdot \frac{1 + \omega \cdot [A]/K_A}{1 + [A]/K_A}$$

where the approximation above is justified by the fact that typical core promoters are very weak such that $[RNAP]/K_{RNAP} \ll 1$ (22-24). The rate of protein expression, $G \propto P$, can then be written as

$$G \approx G_0 \cdot \frac{1 + \omega \cdot [A]/K_A}{1 + [A]/K_A}. \tag{S7}$$

where $G_0$ is a basal protein expression rate, involving the rate of transcription, translation as well as mRNA turnover, in the absence of any regulatory proteins. It is not a number computable from the thermodynamic model; on the other hand, the knowledge of this overall prefactor is not necessary for characterizing the relative effect of transcriptional control.

In the case of a repressor protein *R* which binds to its operator with dissociation constant $K_R$ and precludes RNAP binding at the core promoter, the transcription rate can be written as

$$P = \frac{[RNAP]/K_{RNAP}}{1 + [R]/K_R + [RNAP]/K_{RNAP}} \simeq \frac{[RNAP]}{K_{RNAP}} \cdot \frac{1}{1 + [R]/K_R}$$

and the rate of protein expression as

$$G \approx G_0 \cdot \frac{1}{1 + [R]/K_R}. \tag{S8}$$

For a promoter which is controlled by both a repressor and an activator, the simplest case would be that the two exert their influence on the RNAP independently and do not interact with each other. In this case, the combined response would simply be the *product* of the two expressions given above, i.e.

$$G \approx G_0 \cdot \left( \frac{1+\omega \cdot [A]/K_A}{1+[A]/K_A} \right) \cdot \left( \frac{1}{1+[R]/K_R} \right) \tag{S9}$$

Below, we apply the general thermodynamic to the *lac* promoter.

**c. Activation by cAMP-CRP.** We first address activation by cAMP-CRP, whose concentration is $[CRP^*]$ as given by Eq. (S3). cAMP-CRP binds to the operator site at position -61.5 (see Fig. S6) and recruits the RNA polymerase to the core promoter. In the absence of LacR, this is just the case of "simple activation" considered in Eq. (S7). Denoting the *in vivo* binding constant (25) of cAMP-CRP to the operator site by $K_{CRP}$, and the cooperativity factor of CRP-RNAp interaction by $\omega$, then the rate of LacZ synthesis is

$$G = G_0 \cdot \frac{1+\omega \cdot [CRP^*]/K_{CRP}}{1+[CRP^*]/K_{CRP}}. \tag{S10}$$

where $G_0$ is the same basal rate introduced in Eq. (S7). Eq. (S10) describes the rate of LacZ synthesis. What we measure is the enzymatic activity (*A*) of LacZ (see Methods and Section S1 of Supp Mat) in the steady state of balanced exponential growth. We take *A* to be proportional to the steady-state LacZ concentration $[LacZ]^*$. The latter is given through the kinetic equation

$$\frac{d}{dt}[LacZ] = G - \lambda \cdot [LacZ]$$

with the steady-state solution $G = \lambda \cdot [LacZ]^* \propto \lambda \cdot A$, where λ is the cell growth rate. In our study, we defined the promoter activity to be $\alpha \equiv A \cdot \lambda_{\frac{1}{2}}$ where $\lambda_{\frac{1}{2}} = \lambda / \ln 2$ is the doubling rate of the culture; see Eq. (S1). Hence, $G \propto \alpha$ as expected. Inserting the cAMP-dependence of $[CRP^*]$ as given by Eq. (S3), we have

$$\alpha_{cAMP} = b_0 \cdot \frac{1+\omega \cdot [cAMP]/\kappa_{cAMP}}{1+[cAMP]/\kappa_{cAMP}} \tag{S11}$$

where $b_0 \propto G_0 \cdot \lambda$ and

$$\kappa_{cAMP} = K_{cAMP} \cdot K_{CRP} / [CRP] \cdot \gamma_{cAMP} . \quad \text{(S12)}$$

Eq. (S11) is the Hill form used in the main text for the cAMP response (Eq. 2), with $f_{cAMP} = \omega$, and $C_{cAMP} = \kappa_{cAMP}$. The predicted range for $C_{cAMP}$ listed in Table 2 of the main text is obtained by using Eq. (S12) with $K_{cAMP} = 10\ \mu\text{M}$ (10), $K_{CRP} = 5 - 50$ nM (10), $\gamma_{cAMP} = 0.001$ (see discussion above), and $[CRP] = 1500$ nM (26).

**d. Simple repression by LacR.** Let us first consider the case of "simple repression" by the operator O1 alone (see Fig. S6), without the involvement of DNA looping and without activation by cAMP-CRP. In this case, repression is due solely to the direct steric interference between the binding of LacR to O1 and RNAp to the core promoter (17, 27, 28). Specifically, both activated species LacR* and LacR** can bind to O1 and block the promoter. Denoting the *in vivo* binding constant between either activated species and O1 by $K_1$, the rate of LacZ synthesis is given by Eq. (S8) with $[R = [LacR^*] + 2[LacR^{**}]$, with the factor of 2 in the second term arising from the fact that there are two ways for LacR** to bind to O1. Thus,

$$G = G_0 \Big/ \left(1 + [LacR^*] / K_1 + 2\ [LacR^{**}] / K_1\right), \quad \text{(S13)}$$

Inserting into Eq. (S13) the IPTG dependences of $[LacR^{**}]$ and $[LacR^*]$ as given by Eqs. (S4) and (S5), we obtain the result

$$G = G_0 \Bigg/ \left(1 + \frac{2\ [LacR] / K_1}{\left(1 + [IPTG] / K_{IPTG}\right)^2}\right). \quad \text{(S14)}$$

Note that the factor $2[LacR] / \left(1 + [IPTG] / K_{IPTG}\right)^2$ in the denominator of Eq. (S14) simply gives the concentration of the *dimeric units* of LacR tetramers which are not bound by IPTG.

The value of the factor $2\ [LacR] / K_1$ (denoted by $R$ in the main text) can be inferred by noting that the degree of repression *f*, defined as the ratio of LacZ expression for cells grown in medium with saturating and zero IPTG, is simply given by $1 + R$ accord to Eq. (S14). Oehler et al reported in a 1990 paper (29) that the degree of repression was ~20 for a mutant *lac* promoter containing only the operator O1. In a later paper (30), they reported degrees of repression of 200 and 4700 for the same mutant *lac* promoter in cells expressing 5x and 90x respectively of the amount of Lac tetramers of wildtype cells. These results yield an estimate of $R \approx 20$ or $R \approx 50$ for wildtype cells. In the following analysis we use $R \approx 50$, although results of similar quality are obtained for $R \approx 20$.

**e. LacR-mediated DNA looping.** To include repression due to DNA looping, we take into account of the additional possibility that a completely uninduced LacR tetramer (i.e., LacR**) may bind to two operator sites, the main operator O1 and one of the two auxiliary operator sites O2 or O3; see Fig. S6. This is incorporated into the thermodynamic model by adding two looping terms[3] to the denominator of the expression in Eq. (S13), i.e.,

$$G = G_0 \bigg/ \left( 1 + \frac{[LacR^*]}{K_1} + \frac{2\,[LacR^{**}]}{K_1} + \frac{[LacR^{**}][L_{12}]}{K_1 K_2} + \frac{[LacR^{**}][L_{13}]}{K_1 K_3} \right) \qquad \text{(S15)}$$

with $K_2$ and $K_3$ being the dissociation constant of the LacR tetramer with O2 and O3 respectively, and $[L_{12}]$ and $[L_{13}]$ describing the effective "local concentration" of the repressor at O2 and O3 respectively given that the tetramer is bound at O1; see (22-24). The effective concentrations $[L_{12}]$ and $[L_{13}]$ involve the free energy cost of DNA looping and have not been directly characterized[4] by biochemical or biophysical methods for the *lac* promoter in the absence of CRP binding. Inserting again the IPTG-dependence of $[LacR^{**}]$, we obtain

$$G = G_0 \Bigg/ \left( 1 + \frac{2\,[LacR]/K_1}{\left(1 + [IPTG]/K_{IPTG}\right)^2} + \frac{\left([LacR]/K_1\right)\cdot\left([L_{12}]/K_2 + [L_{13}]/K_3\right)}{\left(1 + [IPTG]/K_{IPTG}\right)^4} \right). \qquad \text{(S16)}$$

As the P*lac* activity $\alpha$ is proportional to $G$ (see above), we have

$$\alpha_{IPTG} = b_0 \Bigg/ \left( 1 + \frac{R}{\left(1 + [IPTG]/K_{IPTG}\right)^2} + \frac{R \cdot L_0}{\left(1 + [IPTG]/K_{IPTG}\right)^4} \right) \qquad \text{(S17)}$$

---

[3] In the full thermodynamic treatment, each of the two looping term, $[LacR^{**}][L_{1i}]/K_1K_i$, where $i$ = 2 or 3, is further divided by the factor $(1+[LacR^*]/K_i + 2\,[LacR^{**}]/K_i)$. This factor takes into account of the fact that looping of $O_1$ with the auxiliary operator $O_i$ is not possible if $O_i$ is occupied itself (referred to as the "squelching" effect. We neglected this effect here as the LacR concentration is sufficiently low in wildtype cells. [Note: the squelching term for O1-O3 looping is completely negligible as $2[LacR] \ll K_3$. O1-O2 looping is actually negligible only for $[IPTG] > K_{IPTG}$. This is indeed the regime where most of the data were taken, i.e., most of the data in Fig. 2a (squares) are larger than $K_{IPTG} \sim 10$ µM. However, the fact that the data point for [IPTG]=0 also falls on the same curve (bottom red) suggests that $[L_{12}]/2K_2 \ll [L_{13}]/2K_3$.]

[4] Their values can be indirectly inferred from the experiments of Oehler et al (30) and Muller et al (37) as was done by Vilar & Leibler (40). However the estimates from the two experiments differ by a factor of 5 or more.

as quoted in Eq. (3) of the main text for the IPTG response, with $R = 2\ [LacR]/K_1$ and $L_0 = \frac{1}{2}\left([L_{12}]/K_2 + [L_{13}]/K_3\right)$. With DNA looping, the degree of repression becomes

$$f = 1 + R \cdot (L_0 + 1)\,. \qquad \text{(S18)}$$

From Eq. (S17), it is also straightforward to compute the dependence of sensitivity (defined e.g., as the maximum logarithmic derivative of the transition region) on the loop parameter $L_0$. We find a monotonic function increasing from ~1.2 for small values of $L_0$ to ~2.5 at $L_0$ = 50.

In a recent study, Oehler et al (31) characterized the IPTG response for Lac systems incapable of DNA looping (either by the removal of the auxiliary operator O3 or by using mutant LacR incapable of tetramerization) resulted in a noticeable decrease in the cooperativity of the IPTG response of the lac promoter. We analyzed the data published in (31) quantitatively and found those results to match very well with the expected behavior developed in this work: As shown in Fig. S5(a), the IPTG responses for the two systems incapable of DNA looping (red triangles and blue squares) are well-fitted by Eq. (S17) with the looping term $L_0$ set to zero (solid line). The two fitting parameters $R$ ~ 200 and $K_{IPTG}$ ~ 6 μM obtained are consistent with that expected for strains overproducing LacR 5-10x. The ~1000x fold change of the wildtype strain is a consequence of residual affinity of overexpressed LacR for the operator sequences (32); a previous study by Oehler in which the fold change is determined by LacR+/- comparison shows an 8000x fold change as predicted by the model (30).

**f. Combinatorial control by LacR and cAMP-CRP.** We now consider the combined control of the *lac* promoter by both LacR and CRP. According to the thermodynamic model, if there is no interaction between CRP-cAMP mediated activation and LacR mediated repression, then the joint dependence would simply be the algebraic *product* of the activation and repression functions (Eqs. (S10) and (S15)) in a way analogous to Eq. (S9), i.e.,

$$G = G_0 \cdot \left( \frac{1 + \omega \cdot [CRP^*]/K_{CRP}}{1 + [CRP^*]/K_{CRP}} \right) \Bigg/ \left( 1 + \frac{[LacR^*]}{K_1} + \frac{2\ [LacR^{**}]}{K_1} + \frac{[LacR^{**}][L_{12}]}{K_1 K_2} + \frac{[LacR^{**}][L_{13}]}{K_1 K_3} \right) \qquad \text{(S19)}$$

Alternatively, the co-dependence of the promoter activity on cAMP and IPTG can be obtained from the product of Eq. (S11) and Eq. (S17),

$$\alpha = b_0 \cdot \left( \frac{1+\omega\cdot[cAMP]/\kappa_{cAMP}}{1+[cAMP]/\kappa_{cAMP}} \right) \Bigg/ \left( 1 + \frac{R}{\left(1+[IPTG]/K_{IPTG}\right)^2} + \frac{R\cdot L_0}{\left(1+[IPTG]/K_{IPTG}\right)^4} \right). \quad \text{(S20)}$$

Note that Eq. (S20) is of the same form as the IPTG response of Eq. (S17), but with the basal activity $b_0$ replaced by $\alpha_{cAMP}$ of Eq. (S11). This corresponds to a simple cAMP-dependent vertical shift of the IPTG response in the log-log plot.

As the data shown in Fig. 2b clearly deviated from the above product form (see below), we examined the consequence of a previously proposed interaction between CRP and DNA looping (18, 33-35). As proposed by Lewis et al (18), upon binding to its operator, CRP induces a 90-130$^o$ bend in the DNA, bringing O1 and O3 into closer proximity. This is thought to increase the local concentration $[L_{13}]$, thereby increasing the strength of the looping interaction between them. Experimental evidence for such CRP enhancement of O1-O3 looping has been found *in vitro* (33-35), but no supporting evidence has been reported so far *in vivo*.

This effect can be incorporated into the thermodynamic model developed above simply by including a cooperativity factor $\Omega_0 > 1$ into the term in the denominator of Eq. (S19) involving the product of $[CRP^*]$ and $[L_{13}]$. This amounts to changing the factor $[L_{13}]$ appearing in Eq. (S19) to

$$[L_{13}]\cdot\left(1+\Omega_0\cdot[CRP^*]/K_{CRP}\right)\Big/\left(1+[CRP^*]/K_{CRP}\right). \quad \text{(S21)}$$

It then follows that the cAMP and IPTG dependence of the promoter activity can be written in the same form as Eq. (S20), but with the looping factor $L_0$ changed to

$$L = L_0 \cdot \frac{1+\Omega\cdot[\text{cAMP}]/\kappa_{cAMP}}{1+[\text{cAMP}]/\kappa_{cAMP}} \quad \text{(S22)}$$

where $\Omega$ is an effective cooperativity factor given by

$$\Omega = \left( \frac{[L_{12}]}{K_2} + \Omega_0 \cdot \frac{[L_{13}]}{K_3} \right) \Bigg/ \left( \frac{[L_{12}]}{K_2} + \frac{[L_{13}]}{K_3} \right). \quad \text{(S23)}$$

The parameters $[L_{12}]$, $[L_{13}]$, and $\Omega_0$ have not been determined from direct biochemical studies. However, Fried et al (33, 35) have measured an effective cooperativity factor defined as the ratio of concentration of the cAMP-CRP-LacR-DNA ternary product with the product of the concentrations of cAMP-CRP-DNA and LacR-DNA binary products. In terms of the parameters defined above, this effective cooperativity factor can be written as

$$\frac{\left(1+\frac{[CRP^*]}{K_{CRP}}\right)\cdot\left(1+R+R\cdot L\right)+\left(\Omega_0-1\right)\cdot\frac{[CRP^*]}{K_{CRP}}R\cdot\frac{[L_{13}]}{2K_3}}{\left(1+\frac{[CRP^*]}{K_{CRP}}\right)\cdot\left(1+R+R\cdot L\right)}. \tag{S24}$$

For saturating amounts of cAMP-CRP and LacR such that $[CRP^*]\gg K_{CRP}$ and $R\gg 1$, the expression in Eq. (S24) is identical to the right hand side of Eq. (S23). Hence, the effective cooperativity factor Ω introduced in Eq. (S22) is just the cooperativity factor of the range 4 – 12 determined experimentally by Fried et al (33, 35).

Finally, we consider the case where $[CRP^*]$ depends nonlinearly on the cAMP concentration of the medium such that the observed cAMP response is described by the general Hill form

$$\alpha_{cAMP}=b_0\cdot\frac{1+\omega\cdot\left([cAMP]/C_{cAMP}\right)^m}{1+\left([cAMP]/C_{cAMP}\right)^m} \tag{S25}$$

with Hill coefficient $m>1$. This is for example the situation for TK250 cells bearing the *cpdA* gene (see blue symbols in Fig. 1b of the main text). Working backward from Eq. (S10), we have

$$[CRP^*]=K_{CRP}\left([cAMP]/C_{cAMP}\right)^m \tag{S26}$$

Using Eq. (S26) instead of Eq. (S3) in the combinatorial control function (Eq. (S19) with $[L_{13}]$ replaced by the expression in (S21), we obtain the following for the co-dependence of the promoter activity:

$$\alpha=b_0\cdot\left(\frac{1+\omega\cdot\left([cAMP]/C_{cAMP}\right)^m}{1+\left([cAMP]/C_{cAMP}\right)^m}\right)\Bigg/\left(1+\frac{R}{\left(1+[IPTG]/K_{IPTG}\right)^2}+\frac{R\cdot L}{\left(1+[IPTG]/K_{IPTG}\right)^4}\right) \tag{S27}$$

with the loop parameter $L$ in Eq. (S27) being

$$L=L_0\cdot\frac{1+\Omega\cdot\left([\mathrm{cAMP}]/C_{cAMP}\right)^m}{1+\left([\mathrm{cAMP}]/C_{cAMP}\right)^m}, \tag{S28}$$

and the cooperativity factor Ω in Eq. (S28) still given by Eq. (S23).

**g. Comparison to alternative models.** While our model described by Eqs. (S20) and (S22) appears to accurately describe the data (See Fig. 2b), one might question whether conceptually simpler models (e.g., generalized Hill functions) might explain the data equally well. Below, we construct several such models and fit to the data surface from strain TK310 using non-linear least squares minimization in Matlab. The points at [IPTG] = 0 are weighted to increase their significance (to simulate the wide range (in log scale) of low IPTG concentrations (< 5 μM) over which the baseline expression from the promoter does not significantly increase). For each model the specified parameters are allowed to vary freely to obtain the optimum fit.

(a) A naïve 6-parameter model based on generic Hill functions and assuming no interaction between LacR and CRP:

$$\alpha = b_0 \cdot \left( \frac{1+\omega\left([cAMP]/C_{cAMP}\right)^{m_{cAMP}}}{1+\left([cAMP]/C_{cAMP}\right)^{m_{cAMP}}} \right) \Bigg/ \left( 1+\frac{R}{1+\left([IPTG]/C_{IPTG}\right)^{m_{IPTG}}} \right) \qquad \text{(S29)}$$

As can be seen from Fig. S8(a), neither the fold change nor the sensitivity to IPTG response is properly captured by this model for the range of cAMP concentrations studied.

(b) A 7-parameter model in which the coefficient of the IPTG dependent term, i.e., *R* in (S29) above, is a four-parameter cAMP-dependent Hill function,

$$R = R_0 \cdot \frac{1+\omega'([cAMP]/C'_{cAMP})^{m_{cAMP}}}{1+([cAMP]/C'_{cAMP})^{m_{cAMP}}} \qquad \text{(S30)}$$

where we assume $C'_{cAMP} = C_{cAMP}$, and $m'_{cAMP} = m_{cAMP}$, but allow $R_0$ and $\omega'$ to vary freely. From Fig. S8(b), we see that the fold change of the IPTG response is better reproduced than model (a), but the overall correspondence between data and model prediction is still poor, particularly for intermediate IPTG concentrations (10-500 μM) for both the high and low cAMP levels.

(c) The same model as (b), but with. $C'_{cAMP}$, and $m'_{cAMP}$ both varying freely. From Fig. S8(c) it is seen that the fit to the data is further improved. However the number of parameters is increased to nine while the fit to data points at intermediate IPTG concentrations is still problematic, especially at low cAMP levels.

(d) The model Eq. (S29) with the modification to $R$ made in Eq. (S30), and with the Hill coefficient $m_{IPTG}$ in (S29) also cAMP dependent. The latter is specified by another Hill function

$$m_{IPTG} = m_0 \cdot \frac{1+\omega''([cAMP]/C''_{cAMP})^{m''_{cAMP}}}{1+([cAMP]/C''_{cAMP})^{m''_{cAMP}}}. \qquad \text{(S31)}$$

Here, we use $C''_{cAMP} = C'_{cAMP} = C_{cAMP}$, $m''_{cAMP} = m'_{cAMP} = m_{cAMP}$ but allow both $\omega''$ and $\omega'$ to vary freely. This 8-parameter model is able to describe the full data set well; see Figure S8(d). This finding is not surprising given the result in Fig. S3(a), that the effective Hill coefficient of the IPTG response is strongly cAMP dependent.

We see that while both our model (i.e. Eqs. (S20) and (S22)) and the model (d) above (i.e., Eqn (S29) (S30) and (S31)) adequately describe the experimental data, our model requires seven parameters while model (d) requires eight. More importantly, the form used in our model is well founded on known biochemical facts as described above, and the values of fitted parameters compare favorably with the known range in all cases when comparison can be made. (Note: The loop cost $L_0$ is the only parameter where independent estimate is not available.) On the other hand, it is difficult to justify biochemically the cAMP-dependence of the Hill coefficient in model (d). The above results therefore demonstrate that our model based on the thermodynamic treatment of DNA looping and its interaction with CRP is the minimal model describing the activity of the *lac* promoter.

# DATA ANALYSIS

For each strain and medium used, the promoter activity $\alpha$ was computed from the raw data (i.e., β-galactosidase activity and growth curves) according to Eq. (S1). Its dependence on either the concentrations of cAMP or IPTG in the medium was then fitted to an effective Hill form

$$\alpha_X = b_X \cdot \frac{1+f_X\left([X]/C_X\right)^{m_X}}{1+\left([X]/C_X\right)^{m_X}} \qquad \text{(S32)}$$

where $[X]$ refers to the concentration of the inducer X (cAMP or IPTG) in the medium. In the fits, the basal activity $b_X$ was directly fixed by the promoter activity obtained for $[X]=0$. The other 3 Hill parameters were obtained by fitting to Eq. (S32) using nonlinear least-square minimization (Matlab 7). Reported errors are the 68.3% (1 standard deviation) confidence intervals for the given parameters.

Because DNA looping is involved in repression of the *lac* promoter by the Lac tetramer (17, 29, 30, 36, 37), the IPTG responses of strains TK230 and TK310 (in medium with no cAMP added) were also fitted to the form (S17) predicted by the DNA looping model above. In the fit, we fixed the value of $R$ to 50 as discussed above. We then used the experimentally determined promoter activity in saturating (1mM) IPTG condition to fix the coefficient $b_0$, and used the remaining data to fit the two parameters $K_{IPTG}$, and $L_0$ according the procedure described above. The results are listed in Table 3 (rows 1 and 2). [The values of the IPTG-LacR dissociation constant $K_{\mathrm{IPTG}}$ obtained are comparable to the results of *in vitro* biochemical studies: In the absence of DNA, the LacR-IPTG interaction was characterized by $K_{\mathrm{IPTG}}$ = 0.8 – 4 μM at physiologically relevant pH values (32, 38, 39). In the presence of a saturating amount of operator DNA however, a much weaker LacR-IPTG affinity characterized by $K_{\mathrm{IPTG}} \approx 80$ μM was found, although with non-specific DNA fragments the smaller value of $K_{\mathrm{IPTG}}$ was obtained as before (20). Our results with $K_{\mathrm{IPTG}}$ = 15 ~ 20 μM lie in between this range, presumably reflecting the interaction of LacR with both specific and non-specific sequences *in vivo*.]

For the IPTG response of strain TK310 and TK250 grown in medium with various cAMP concentrations (the data in Fig. 2b and Fig. S7b), we determined the looping parameter $L$ for the IPTG dependence at each cAMP concentration by first computing the degree of repression $f$ as the ratio of $b_0$ (determined as above) and the value of the promoter activity at [IPTG] = 0, and then extracting the loop parameter from $f$ using Eq. (S18). The values of $L$ obtained were plotted in Fig. 2a and Fig. S7a for strain TK310 and TK250 respectively. Then for the IPTG response of strain TK310, the predicted form (S20) was plotted as the dashed line in Fig. 2b using the values of $L$ shown in Fig. 2a and other parameter values as listed in Table 2 (row 2) and Table 3 (row 2). Similarly, for the IPTG response of strain TK250, the predicted form (S27) was plotted as the dashed line in Fig. S7b using the values of $L$ shown in Fig. S7a and other parameter values as listed in Table 2 (row 1) and Table 3 (row 3).

Finally, the cAMP dependence of the loop parameter $L$ was analyzed by fitting their values to the expected form according to the extended thermodynamic model, Eq. (S22) for strain TK310 and Eq. (S28) for strain TK250. In these fits, we varied only the effective cooperativity factor Ω, taking the values of the other parameters from Table 2 (row 2) and Table 3 (row 2) for strain TK310, and Table 2 (row 1) and Table 3 (row 3) for strain TK250. Estimate of error on Ω was again obtained using the procedure described above.